\documentclass{article}
\usepackage{graphicx} 
\usepackage[utf8]{inputenc}
\usepackage[margin=1in]{geometry}
\usepackage[super, sort&compress]{natbib}
\usepackage{placeins}
\usepackage{amsmath}
\usepackage{multirow}
\usepackage{xcolor}
\usepackage{mathtools}
\usepackage{setspace}
\usepackage{lscape}
\usepackage{tikz}
\usepackage{float}
\usepackage{rotating}
\usepackage{url}

\allowdisplaybreaks

\title{Bayesian kernel machine regression for heteroscedastic health outcome data}
\author{Melissa J. Smith, Ihsan E. Buker, Kristina M. Zierold, Lonnie Sears, Cassandra Newsom,\\ Wilco Zijlmans, Maureen Lichtveld, Jeffrey K. Wickliffe}

\date{}

\begin{document}

\maketitle
\section*{Abstract}
The field of environmental epidemiology has placed an increasing emphasis on understanding the health effects of mixtures of metals, chemicals, and pollutants in recent years. Bayesian Kernel Machine Regression (BKMR) is a statistical method that has gained significant traction in environmental mixture studies due to its ability to account for complex non-linear relationships between the exposures and health outcome and its ability to identify interaction effects between the exposures. However, BKMR makes the crucial assumption that the error terms have a constant variance, and this assumption is not typically checked in practice. In this paper, we create a diagnostic function for checking this constant variance assumption in practice and develop Heteroscedastic BKMR (HBKMR) for environmental mixture analyses where this assumption is not met. By specifying a Bayesian hierarchical variance model for the error term variance parameters, HBKMR produces updated estimates of the environmental mixture's health effects and their corresponding 95\% credible intervals. We apply HBKMR in two real-world case studies that motivated this work: 1) Examining the effects of prenatal metal exposures on behavioral problems in toddlers living in Suriname and 2) Assessing the impacts of metal exposures on simple reaction time in children living near coal-fired power plants in Kentucky. In both case studies, HBKMR provides a substantial improvement in model fit compared to BKMR, with differences in some of the mixture effect estimates and typically narrower 95\% credible intervals after accounting for the heteroscedasticity.
\section{Introduction}
The field of environmental epidemiology has placed an increasing emphasis on understanding the health effects of mixtures of metals, chemicals, and pollutants in recent years. Traditionally, much of the literature on environmental exposures has assessed the impact of one environmental exposure at a time due to the dearth of statistical methods available for environmental mixtures. This ``one-at-a-time'' approach can be problematic because it disregards correlations between environmental exposures, and it does not capture the complex joint relationships that these exposures often have with health outcomes of interest. To support the development of statistical methods more suitable for environmental mixtures, the National Institute of Environmental Health Sciences (NIEHS) launched the Powering Research through Innovative Methods for Mixtures in Epidemiology (PRIME) program in 2017.\cite{prime} A method that garnered significant attention during this program, and that continues to gain popularity in the field, is Bayesian Kernel Machine Regression (BKMR).\cite{bobb2015bayesian,bobb2018statistical}

BKMR is a flexible statistical method with several advantages for environmental mixture modeling. First, it enables researchers to model the effects of each exposure on the outcome in a nonlinear manner, capturing complex dose-response relationships that are ubiquitous in environmental studies. Additionally, BKMR can identify interaction effects between exposures without the need to pre-specify potential interactions. By employing a spike-and-slab prior distribution, BKMR incorporates a built-in variable selection mechanism. This feature aids in identifying the most influential components of the mixture while minimizing the impact of less important components. Moreover, BKMR handles highly correlated environmental exposures effectively through its hierarchical variable selection capabilities. \cite{bobb2015bayesian, bobb2018statistical} To facilitate its use in practice, Bobb et al \cite{bobb2018statistical} developed the \texttt{bkmr} R package, which includes functions for model fitting, visualizing key aspects of the exposure-response function, and generating summaries of mixture effects on the outcome.

Numerous extensions of BKMR have been proposed since its inception. While initially proposed for continuous outcomes, extensions of BKMR for binary outcomes, count outcomes, and other outcome types have been developed more recently. \cite{bobb2018statistical, mutiso2024bayesian, mou2024generalized} BKMR has also been extended for use in studies with multiple time points including studies with longitudinal outcomes and exposures. \cite{liu2018bayesian, wilson2022kernel} Devick et al \cite{devick2022bayesian} developed a BKMR-integrated causal mediation analysis method when the environmental mixture represents the exposure of interest and may affect a mediator variable. While many important extensions of BKMR have been developed, BKMR and its continuous-outcome extensions make the crucial assumption of a constant variance for the model error terms, and this does not always hold in practice. In regression analyses, it is well known that failing to account for heteroscedasticity can affect inferences on the model coefficients, leading to a decrease in power or an increase in type I errors.\cite{rosopa2013managing, pelenis2014bayesian} Consequently, heteroscedasticity within the BKMR model has the potential to impact inferences surrounding the mixture function cross
sections and the conclusions about the effect of the environmental mixture on the outcome.

In this paper, we propose heteroscedastic BKMR, or HBKMR, an extension of BKMR developed for datasets with heteroscedasticity. This method allows researchers to draw reliable inferences on the health effects of environmental mixtures under violations of the constant variance of error terms assumption. We also develop a BKMR diagnostic function that can be used to assess the homoscedasticity assumption in BKMR model fits. The remainder of this paper is organized as follows. In Section \ref{sec:existing}, we describe the BKMR method, relevant notation, and an approach for detecting heteroscedasticity with BKMR. In Section \ref{sec:hbkmr}, we introduce the HBKMR method and highlight its flexible implementation in the NIMBLE statistical software. We illustrate the application of HBKMR through two real-world case studies in Section \ref{sec:comparison} and compare its performance to BKMR in these settings. Finally, Section \ref{sec:discussion} includes a discussion of the strengths and limitations of our method and considerations for its use in practice. 

\section{Bayesian Kernel Machine Regression} \label{sec:existing}
Let $Y_i$ denote a continuous health outcome of interest for individual $i$, where $i = 1, \dots, N$ and let $\boldsymbol{y}$ denote the entire vector of outcomes. Assuming there are $M$ exposures in the mixture, where $m = 1, \dots, M$, let  $\boldsymbol{z}_i = \left(z_{i,1}, \dots, z_{i,M}\right)'$ denote the vector of exposure values for individual $i$. We express the full $N\times M$ matrix of exposures as $\boldsymbol{Z}$. Finally, let $\boldsymbol{x}_i$ represent a length-$P$ vector of important covariates for individual $i$, and let $\boldsymbol{X}$ denote the $N\times P$ matrix of all individuals' covariates. The BKMR model for continuous outcomes is specified as: 
\begin{align*}
Y_i= h(\boldsymbol{z}_i )+ \boldsymbol{x}_i'\boldsymbol{\beta}+ \epsilon_i \hspace{0.2cm} \text{where } \epsilon_i  \overset{ind}{\sim} \text{Normal}(0,\sigma^2).
\end{align*}
In this model formulation, $h(\cdot)$ represents a flexible function of the set of exposures, which is obtained using a Gaussian kernel. This is the component of the BKMR model on which inferences are typically made.  The BKMR model also allows researchers to adjust for important covariates through the model term $\boldsymbol{x}_i'\boldsymbol{\beta}$.  

Let $h_i = h(\boldsymbol{z}_i)$ and $\boldsymbol{h} = (h_1, \dots, h_N)'$. The flexible mixture function is modeled as:

\begin{align*}
\boldsymbol{h} \mid \tau, \boldsymbol{r}, \boldsymbol{Z} &\sim \operatorname{MVN}\left(\boldsymbol{0}, \tau \boldsymbol{K}_{\boldsymbol{r}}\right)
\end{align*}
where the $i,j$th entry of $\boldsymbol{K}_{\boldsymbol{r}}$ is:

$$K(z_i, z_j) = \exp\left(-\sum_{m=1}^Mr_m(z_{im}-z_{jm})^2\right).$$

This kernel naturally shrinks the health effects for individuals who have similar exposure profiles toward one another. Furthermore, it captures interactive effects and non-linearities between the exposures and the health outcome. If variable selection is of interest, a component-wise or hierarchical spike-and-slab prior may be placed on $\boldsymbol{r}$. In this paper, we focus on the setting in which the retention of all exposures in the mixture is of greatest interest to more accurately estimate the joint effects of the mixture.

To implement BKMR, the authors developed the \texttt{bkmr} R package.\cite{bobb2018statistical} This package employs a Gibbs sampler algorithm with Metropolis-Hastings steps for drawing posterior samples of $\boldsymbol{r}$ and  $\lambda = \frac{\tau}{\sigma^2}$. After fitting the model, the package enables researchers to compute and visualize important cross sections of the $h(\cdot)$ function to identify dose-response relationships between the metals and the outcome, quantify the joint effect of the metal mixture, and uncover pairwise interactions between metals. For example, one cross section that may be used to capture the joint effect of the collection of exposures on the health outcome is
$h(\boldsymbol{z}_{0.75}) - h(\boldsymbol{z}_{0.50})$, which reflects the average change in health outcome (adjusted for covariates), when changing all exposures in the mixture from their 50th percentile values to their 75th percentile values.

While the \texttt{bkmr} R package has propelled the method's use in environmental epidemiology research, the posterior sampling algorithm and current implementation provides limited flexibility for changes to the model specification. In this paper, we therefore implement BKMR in NIMBLE,\cite{nimble} allowing for a fair comparison to our proposed HBKMR method, while enabling more flexibility in prior distributions and model modifications. Notably, our BKMR implementation in NIMBLE results in faster convergence and more efficient sampling of the $\boldsymbol{r}$ parameters without the need to manually tune the proposal distributions in the Metropolis-Hastings steps. Section \ref{sec:hbkmr} provides a detailed description of prior distributions used in both our BKMR and HBKMR implementations.
\subsection{Detecting heteroscedasticity in BKMR models}
As previously described, BKMR makes the important assumption of a constant error term variance, $\sigma^2$. This assumption is not always realistic, yet it is not typically checked in practice. We therefore develop a diagnostic function for BKMR models, employing a definition of Bayesian residuals described by Li and Wulff.\cite{li2018bayesian}

Let $\boldsymbol{e} = \left(e_{1}, \dots e_{n}\right)'$ represent a vector of Bayesian residuals from the BKMR model fit, where the residuals are defined as:

$$e_i = Y_i - E(h(\boldsymbol{z}_i) + \boldsymbol{x}_i'\boldsymbol{\beta} \mid \boldsymbol{y}) = Y_i - E(h(\boldsymbol{z}_i)\mid \boldsymbol{y}) - E(\boldsymbol{x}_i'\boldsymbol{\beta} \mid \boldsymbol{y}).$$ This definition of Bayesian residuals relies on posterior means, rather than calculations based on predictions of $\boldsymbol{y}$ that employ the entire posterior distribution of $h(\boldsymbol{z}_i)$, so they can be efficiently computed from a fitted BKMR model. We utilize the approximation for the mean of $h(z_i)$ proposed by Bobb et al\cite{bobb2018statistical}, which involves estimating $E(h(z_i)\mid \boldsymbol{y})$ from the distribution of $\boldsymbol{h}\mid \boldsymbol{y}$ that is parametrized by the posterior mean of $\tau$ and the posterior mean of $\boldsymbol{r}$. We also propose approximating residuals using a linear regression model with main effects, pairwise interaction terms for all combinations of exposures, and main effects for important covariates for faster computation. Under each residual definition, our function produces residual versus fitted plots and residual versus predictor plots, where Bayesian fitted values are defined as $E(h(\boldsymbol{z}_i) + \boldsymbol{x}_i'\boldsymbol{\beta} \mid \boldsymbol{y})$, to examine patterns in the residual variance. The predictors in these plots may be any variable in $\boldsymbol{X}$ or $\boldsymbol{Z}$.

\section{Heteroscedastic Bayesian Kernel Machine Regression} \label{sec:hbkmr}
We propose Heteroscedastic BKMR (HBKMR) for situations where the BKMR diagnostic plots suggest a non-constant variance of the error terms. The HBKMR model is specified as:
\begin{align*}
Y_i &= h(\boldsymbol{z}_i) + \boldsymbol{x}_i'\boldsymbol{\beta} + \epsilon_i, \hspace{0.2cm} \text{where}\hspace{0.2cm}
 \epsilon_i \overset{ind}{\sim} \operatorname{N}(0, \sigma_i^2) \hspace{0.2cm}\text{and} \hspace{0.2cm}
 \log(\sigma_i^2) = \boldsymbol{w}_i'\boldsymbol{\gamma}.
\end{align*}
In matrix notation, this model can be written as:
\begin{align*}
\boldsymbol{y} &= \boldsymbol{h} + \boldsymbol{X}\boldsymbol{\beta} + \boldsymbol{\epsilon}, \hspace{0.2cm} \text{where}\hspace{0.2cm}
\boldsymbol{\epsilon} \sim MVN\left(\boldsymbol{0}, \boldsymbol{S}_{\gamma}\right).
\end{align*}
In this model specification, $\boldsymbol{S}_{\gamma}$ is an $N\times N$ diagonal matrix with $i$th diagonal entry equal to $\exp\left(\boldsymbol{w}_i'\boldsymbol{\gamma}\right)$, and $h(\cdot)$ is the same function specified in the BKMR model in Section \ref{sec:hbkmr}.

HBKMR incorporates a Bayesian hierarchical variance model into its specification and employs a log link function on $\sigma^2_i$ to constrain the error variances to be strictly positive. A similar approach has been employed in other Bayesian hierarchical modeling contexts to address heteroscedasticity. \cite{parker2021general} Here, $\boldsymbol{w}_i$ is a vector for individual $i$ that includes one or more predictors identified through the BKMR diagnostic plots. If $Q$ predictors are identified for consideration in the variance model, then $\boldsymbol{w}_i = (1, w_{i,1}, \dots, w_{i,Q})'$ to accommodate a model intercept and $\boldsymbol{W}$ is an $N\times (Q+1)$-dimensional matrix. $\boldsymbol{\gamma}'$ is a vector of length $Q+1$ of coefficients corresponding to the predictors in the variance model. Importantly, if $Q = 0$ then $w_i = 1$ for all $i$. This special case is an alternative parametrization of the BKMR model since $\sigma_i^2 = \exp(\gamma_1) = \sigma^2$. Thus, the two models can be directly compared with the same prior distributions using our implementation.

\subsection{Prior and posterior distributions}
To complete the Bayesian model specification, we specify the following prior distributions on the remaining parameters in the HBKMR model:

\begin{itemize}
\item $\beta_p \sim \text{Normal}(0, 1000)$ for  $p = 1, \dots, P$
\item $\gamma_q \sim \text{Normal}(0, 1000)$ for  $q = 1, \dots, Q + 1$
\item $\sqrt{\tau} \sim$ Uniform(0, 100)
\item $r_m \sim$ Inverse-Uniform(0,100) for $m = 1, \dots, M$
\end{itemize}
The Uniform(0,100) distribution on $\sqrt{\tau}$, a global standard deviation term, has been shown to be robust to choice of upper limit while remaining largely non-informative. It is therefore a recommended alternative to an Inverse-Gamma($\epsilon$, $\epsilon$) prior on $\tau$, where $\epsilon$ represents a small positive value such as 0.001.\cite{gelman2006prior} For each $r_m$ in the BKMR model, the Inverse-Uniform(0,100) distribution is a stable choice of prior that is frequently used in practice. Thus, we use this prior when comparing HBKMR and BKMR. However, is important to highlight that our NIMBLE-based code is easily modifiable and other prior distributions may be applied to $\boldsymbol{r}$. The inverse-uniform prior minimizes the risk of overfitting HBKMR to the data and is the default prior for $\boldsymbol{r}$ in the \texttt{bkmr} R package.\cite{bobb2018statistical}

Given the prior distributions and the HBKMR likelihood, the full posterior distribution for HBKMR can be factored as follows:
\begin{align*}
p(\boldsymbol{\beta}, \tau, \boldsymbol{r}, \boldsymbol{h}, \boldsymbol{\gamma}\mid \boldsymbol{y}, \boldsymbol{X}, \boldsymbol{Z}, \boldsymbol{W}) &\propto p(\boldsymbol{y} \mid \boldsymbol{\beta}, \tau, \boldsymbol{r}, \boldsymbol{h}, \boldsymbol{\gamma}, \boldsymbol{X}, \boldsymbol{W})p(\boldsymbol{\beta}, \tau, \boldsymbol{r}, \boldsymbol{h}, \boldsymbol{\gamma}, \boldsymbol{Z}) \\
&\propto p(\boldsymbol{y} \mid \boldsymbol{\beta}, \boldsymbol{h}, \boldsymbol{\gamma}, \boldsymbol{X}, \boldsymbol{W})p(\boldsymbol{h}\mid \tau, \boldsymbol{r}, \boldsymbol{Z})p(\boldsymbol{\beta})p(\boldsymbol{r})p(\tau)p(\boldsymbol{\gamma})
\end{align*}
For computational stability, and following the work of Bobb et al,\cite{bobb2015bayesian, bobb2018statistical} we integrate out $\boldsymbol{h}$ from this joint posterior distribution. This results in a joint posterior distribution of the remaining parameters that can be factored as:
\begin{align*}
p(\boldsymbol{\beta}, \tau, \boldsymbol{r}, \boldsymbol{\gamma}\mid \boldsymbol{y}, \boldsymbol{X}, \boldsymbol{Z}, \boldsymbol{W}) 
&\propto p(\boldsymbol{y} \mid \boldsymbol{\beta},\boldsymbol{\gamma}, \tau, \boldsymbol{r}, \boldsymbol{X}, \boldsymbol{Z}, \boldsymbol{W})p(\boldsymbol{\beta})p(\boldsymbol{r})p(\tau)p(\boldsymbol{\gamma}) \hspace{0.2cm} \text{where}\\
\boldsymbol{y} &\mid \boldsymbol{\beta},\boldsymbol{\gamma}, \tau, \boldsymbol{r}, \boldsymbol{X}, \boldsymbol{Z}, \boldsymbol{W} \sim MVN\left(\boldsymbol{X\beta}, \tau\boldsymbol{K}_{\boldsymbol{r}} + \boldsymbol{S}_{\boldsymbol{\gamma}}\right).
\end{align*}
See the appendix for the full derivation of this posterior distribution. We implement HBKMR in NIMBLE by using the likelihood and priors specified in the final posterior distribution provided above.
\subsection{Inference on the exposure-response function} \label{sec:estimate_exp_resp}
To draw inferences on the effect of the exposure mixture on the health outcome, we compute various cross-sections of the $h(\cdot)$ function. We leverage properties of the joint multivariate normal distribution of $\boldsymbol{y}$ and $\boldsymbol{h}$ and samples from the aforementioned posterior distribution. To do this, we use the joint distribution of $\boldsymbol{y}$ and $\boldsymbol{h}$, which is:
\begin{align*}
\begin{pmatrix}
\boldsymbol{y} \\
\boldsymbol{h}
\end{pmatrix} \sim MVN \left( \boldsymbol{\mu} =
\begin{pmatrix}
\boldsymbol{X\beta} \\
\boldsymbol{0}
\end{pmatrix}, \boldsymbol{\Sigma} =  \begin{pmatrix}
\tau\boldsymbol{K}_{\boldsymbol{r}} + \boldsymbol{S}_{\gamma} & \tau\boldsymbol{K}_{\boldsymbol{r}} \\ \tau\boldsymbol{K}_{\boldsymbol{r}} & \tau\boldsymbol{K}_{\boldsymbol{r}}
\end{pmatrix}\right).
\end{align*}
Given that $\boldsymbol{h}$ represents the vector of values of the exposure-response function at each individual's observed exposure values, we must also obtain posterior predictions of $\boldsymbol{h}$, which we call $\boldsymbol{h}^{new}$, at a new set of exposure values, $\boldsymbol{Z}^{new}$. The set of exposure values in $\boldsymbol{Z}^{new}$ will reflect exposure levels corresponding to different cross-sections of the exposure-response function that we aim to plot. For example, we may want to estimate $h$ when all of the exposures are set to their median concentrations, so the median concentrations would be included in $\boldsymbol{Z}^{new}$. The joint posterior distribution of $\boldsymbol{y}$ and $\boldsymbol{h}^{new}$ is:

\begin{align*}
\begin{pmatrix}
\boldsymbol{y} \\
\boldsymbol{h}^{new}
\end{pmatrix} \sim MVN \left( \boldsymbol{\mu} =
\begin{pmatrix}
\boldsymbol{X\beta} \\
\boldsymbol{0}
\end{pmatrix}, \boldsymbol{\Sigma} =  \begin{pmatrix}
\tau\boldsymbol{K}_{\boldsymbol{r}}(\boldsymbol{Z}, \boldsymbol{Z}) + \boldsymbol{S}_{\gamma}(\boldsymbol{W}) & \tau\boldsymbol{K}_{\boldsymbol{r}}(\boldsymbol{Z}, \boldsymbol{Z}^{new}) \\ \tau\boldsymbol{K}_{\boldsymbol{r}}(\boldsymbol{Z}^{new}, \boldsymbol{Z}) & \tau\boldsymbol{K}_{\boldsymbol{r}}(\boldsymbol{Z}^{new}, \boldsymbol{Z}^{new})
\end{pmatrix}\right)
\end{align*}
where $\boldsymbol{K}_{\boldsymbol{r}}(\boldsymbol{Z}, \boldsymbol{Z}) = \boldsymbol{K}_{\boldsymbol{r}}$ computed under the original exposure values, $\boldsymbol{K}_{\boldsymbol{r}}(\boldsymbol{Z}, \boldsymbol{Z}^{new})$ and $\boldsymbol{K}_{\boldsymbol{r}}(\boldsymbol{Z}^{new}, \boldsymbol{Z})$ are computed with both the original and new exposure values, and $\boldsymbol{K}_{\boldsymbol{r}}(\boldsymbol{Z}^{new}, \boldsymbol{Z}^{new})$ uses only the new exposure values. Similarly, $\boldsymbol{S}_{\gamma} = \boldsymbol{S}_{\gamma}(\boldsymbol{W})$, as the observed $\boldsymbol{W}$ matrix is used to calculate $\boldsymbol{S}_{\gamma}$.  Using properties of the multivariate normal distribution, the following conditional distribution is used to obtain estimates of the exposure-response function under different exposure values:
\begin{align*}
\boldsymbol{h}^{new} \mid \boldsymbol{y} \sim MVN(\boldsymbol{\mu} &=\tau\boldsymbol{K}_{\boldsymbol{r}}(\boldsymbol{Z}^{new},\boldsymbol{Z})\left(\tau\boldsymbol{K}_{\boldsymbol{r}}(\boldsymbol{Z}, \boldsymbol{Z}) + \boldsymbol{S}_{\boldsymbol{\gamma}}(\boldsymbol{W}))^{-1}(\boldsymbol{y - X\beta}\right),\\
\boldsymbol{\Sigma} &= \tau\boldsymbol{K}_{\boldsymbol{r}}(\boldsymbol{Z}^{new},\boldsymbol{Z}^{new}) - \tau^2\boldsymbol{K}_{\boldsymbol{r}}(\boldsymbol{Z}^{new},\boldsymbol{Z})(\tau\boldsymbol{K}_{\boldsymbol{r}}(\boldsymbol{Z}, \boldsymbol{Z}) + \boldsymbol{S}_{\boldsymbol{\gamma}}(\boldsymbol{W}))^{-1}\boldsymbol{K}_{\boldsymbol{r}}(\boldsymbol{Z},\boldsymbol{Z}^{new}))
\end{align*}
Consistent with BKMR, HBKMR estimates $E(\boldsymbol{h}^{new}\mid \boldsymbol{y})$ with $E(\boldsymbol{\mu})$, the mean of the posterior samples of $\boldsymbol{\mu}$, and $Var(E(\boldsymbol{h}^{new}\mid \boldsymbol{y}))$ by taking the sum of $E(\boldsymbol{\Sigma})$ and $Var(\boldsymbol{\mu})$. $E(\boldsymbol{\Sigma})$ corresponds to the mean of the posterior samples of $\boldsymbol{\Sigma}$ and $Var(\boldsymbol{\mu})$ can be found by computing the covariance matrix of the posterior samples of $\boldsymbol{\mu}$. These values are then used to compute 95\% credible intervals for cross-sections of $h$ through a normal approximation to the posterior distribution.
\subsection{Inference on predicted outcomes}
In addition to obtaining estimates of the exposure-response function under heteroscedasticity, we can obtain predicted outcomes and corresponding credible intervals for individuals in the presence of heteroscedasticity using HBKMR.

Let $\boldsymbol{y}^{new}$ represent the predicted health outcome for either an existing individual in the dataset or for a new individual. Similarly, let $\boldsymbol{X}^{new}$ reflect a matrix of new or existing covariate values corresponding to the health outcomes we want to predict, and let $\boldsymbol{Z}^{new}$ reflect the corresponding exposure values. We also define $\boldsymbol{S}_{\gamma}(\boldsymbol{W}^{new})$ to be the covariance matrix for the error terms computed under $\boldsymbol{W}^{new} = (\boldsymbol{X}^{new}, \boldsymbol{Z}^{new})$. The joint distribution of $\boldsymbol{y}$ and $\boldsymbol{y}^{new}$ is the following multivariate normal distribution:
\begin{align*}
\begin{pmatrix}
\boldsymbol{y} \\
\boldsymbol{y}^{new}
\end{pmatrix} \sim MVN \left( \boldsymbol{\mu} =
\begin{pmatrix}
\boldsymbol{X\beta} \\
\boldsymbol{X}^{new}\boldsymbol{\beta}
\end{pmatrix}, \boldsymbol{\Sigma} =  \begin{pmatrix}
\tau\boldsymbol{K}_{\boldsymbol{r}}(\boldsymbol{Z}, \boldsymbol{Z}) + \boldsymbol{S}_{\gamma}(\boldsymbol{W}) & \tau\boldsymbol{K}_{\boldsymbol{r}}(\boldsymbol{Z}, \boldsymbol{Z}^{new}) \\ \tau\boldsymbol{K}_{\boldsymbol{r}}(\boldsymbol{Z}^{new}, \boldsymbol{Z})  & \tau\boldsymbol{K}_{\boldsymbol{r}}(\boldsymbol{Z}^{new}, \boldsymbol{Z}^{new}) + \boldsymbol{S}_{\gamma}(\boldsymbol{W}^{new})
\end{pmatrix}\right)
\end{align*}
 Using properties of the multivariate normal distribution, the following conditional distribution can be used to obtain the posterior predictive distribution of the health outcomes:
\begin{align*}
\boldsymbol{y}^{new} \mid \boldsymbol{y} \sim MVN(\boldsymbol{\mu} &=\boldsymbol{X}^{new}\boldsymbol{\beta} + \tau\boldsymbol{K}_{\boldsymbol{r}}(\boldsymbol{Z}^{new},\boldsymbol{Z})\left(\tau\boldsymbol{K}_{\boldsymbol{r}}(\boldsymbol{Z}, \boldsymbol{Z}) + \boldsymbol{S}_{\boldsymbol{\gamma}}(\boldsymbol{W}))^{-1}(\boldsymbol{y - X\beta}\right),\\
\boldsymbol{\Sigma} &= \tau\boldsymbol{K}_{\boldsymbol{r}}(\boldsymbol{Z}^{new}, \boldsymbol{Z}^{new}) + \boldsymbol{S}_{\gamma}(\boldsymbol{W}^{new}) - \\ &\tau^2\boldsymbol{K}_{\boldsymbol{r}}(\boldsymbol{Z}^{new},\boldsymbol{Z})(\tau\boldsymbol{K}_{\boldsymbol{r}}(\boldsymbol{Z}, \boldsymbol{Z}) + \boldsymbol{S}_{\boldsymbol{\gamma}}(\boldsymbol{W}))^{-1}\boldsymbol{K}_{\boldsymbol{r}}(\boldsymbol{Z},\boldsymbol{Z}^{new}))
\end{align*}
HBKMR estimates the posterior mean and variance of $\boldsymbol{y}^{new}$ using the same approach as in Section \ref{sec:estimate_exp_resp}. This allows for credible intervals for $\boldsymbol{y}^{new}$ to be constructed without needing to draw samples from the entire posterior predictive distribution.
\section{Case Studies} \label{sec:comparison}
In this section, we illustrate the utility of HBKMR and compare it to BKMR through two case studies. All metal concentrations were log-transformed and subsequently centered and scaled prior to inclusion in the BKMR and HBKMR models. In each case study, we drew 80,000 posterior samples from the HBKMR and BKMR model fits after an initial burn-in period of 20,000 posterior draws. We examined trace plots for each parameter to check for convergence to the posterior distribution and computed effective sample sizes for the model parameters to ensure that the 80,000 samples corresponded to a large number of independent samples.

To compare HBKMR and BKMR, we computed the Widely Applicable Information Criterion (WAIC) value for each model fit. \cite{waic} We also created overlayed or side-by-side BKMR and HBKMR plots to compare both point estimates and 95\% credible interval widths of single, pairwise, and joint exposure effects on the health outcomes of interest. Finally, we computed prediction plots under BKMR and HBKMR to illustrate the ability of HBKMR to appropriately quantify uncertainty in predictions.
\subsection{Prenatal metal exposure and behavioral problems in toddlers living in Suriname}
We first applied HBKMR to data from the Caribbean Consortium for Research in Environmental and Occupational Health (CCREOH) Cohort study. CCREOH is an environmental epidemiological cohort of 1,100 mother-child dyads in the Republic of Suriname, South America.\cite{zijlmans2020caribbean} The study assessed prenatal exposures to several metals and elements and evaluated children’s neurodevelopment and behavior in association with these environmental exposures.\cite{wickliffe2021exposure, abdoel2023geographic} Pregnant women were recruited from three regions of Suriname with diverse sources of environmental exposures: (1) Paramaribo, where pesticides are primarily used for residential purposes; (2) Nickerie, the major rice producing region in western Suriname where use of pesticides containing manganese (Mn) and tin (Sn) is abundant; and (3) the tropical rainforest interior, where mercury (Hg) is used in artisanal gold mining and the population is highly dependent on consumption of contaminated fish.

 Behavioral functioning was measured using the parent-reported Child Behavior Checklist (CBCL). The CBCL for ages 1.5–5 (CBCL/1.5–5) is a caregiver-reported instrument used to assess a wide range of emotional and behavioral problems in young children, including key externalizing behaviors such as aggression and hyperactivity. \cite{achenbach2000manual} Comprising 99 items rated on a 3-point scale, the CBCL generates scores across syndrome subscales (e.g. aggressive problems, attention problems, and social problems) and composite domains (e.g. total problems, internalizing problems, and externalizing problems), and demonstrates strong psychometric properties, including good sensitivity and specificity for detecting clinically relevant concerns.\cite{warnick2008screening} Prior studies have linked prenatal exposure to heavy metals with increased behavioral dysregulation as measured by CBCL. For example, Boucher et al\cite{boucher2012prenatal} found that prenatal methylmercury and postnatal Pb exposure were associated with ADHD-related symptoms and behavioral dysregulation in Inuit children. A broader meta-analysis from Nilsen et al\cite{m2020systematic} concluded that prenatal exposure to heavy metals— including lead (Pb), Hg, and Cadmium (Cd) — is consistently associated with externalizing behavior problems such as aggression and conduct difficulties. 

This case study investigated the relationship between prenatal exposure to Pb, Hg, Mn, and Cd and both total problems and aggressive behavioral problems in toddlers in the CCREOH study. In addition to Pb, Hg, and Cd, Mn has been linked to behavioral dysregulation in children, particularly symptoms of hyperactivity and inattention.\cite{rodriguez2013association} Therefore, we included it in the metal mixture alongside Pb, Hg, and Cd.

The analytic sample in this case study consisted of 692 children with complete data on exposures and covariates, including maternal education, geographic district, maternal age, ethnicity, and child’s sex. Among these children, most resided in the Paramaribo district (59.3\%), followed by Nickerie (23.4\%) and the Interior district (17.3\%). In terms of ethnicity, 39.5\% of the children were of African descent, 29.6\% were of Asian descent, and 30.9\% were classified as other, and approximately half of the children were male.  The median number of total problems reported was 30 (IQR: 20–42). The distribution of blood metal concentrations is summarized in Table \ref{tab:suriname_metals}.
Pairwise Spearman correlations among the four metals ranged from –0.24 to 0.50, indicating weak to moderate associations. The strongest correlation was observed between Pb and Hg ($r = 0.50$), while correlations between other pairs, such as Hg and Cd ($r = -0.24$) or Pb and Cd ($r = -0.01$), were relatively weak. Variance inflation factors were also examined, and no evidence of multicollinearity was found in this analysis. To mitigate the right-skewness in CBCL outcomes, behavior counts were square-root transformed. All models were adjusted for the aforementioned covariates.

\begin{table}[h]
\centering
\caption{Quantiles of maternal blood metal concentrations ($\mu g/L$) for the 692 children in Suriname.}
\label{tab:suriname_metals}
\begin{tabular}{|l|l|l|l|l|l|}
\hline
\textbf{\begin{tabular}[c]{@{}l@{}}Maternal blood metal \\ concentration \\ ($\mu g/L$)\end{tabular}} & \textbf{10\%} & \textbf{25\%} & \textbf{50\%} & \textbf{75\%} & \textbf{90\%} \\ \hline
Pb  & 0.89  & 1.20  & 1.79   & 3.09  & 7.20  \\ \hline
Hg  & 1.09  & 1.78  & 2.95   & 5.58  & 13.50 \\ \hline
Mn  & 8.11  & 10.50 & 13.10  & 17.40 & 22.63 \\ \hline
Cd  & 0.09  & 0.13  & 0.19  & 0.28  & 0.37  \\ \hline
\end{tabular}
\end{table}

\subsubsection{Total behavioral problems}

\begin{figure}[H]
\centering
\includegraphics[scale=0.63]{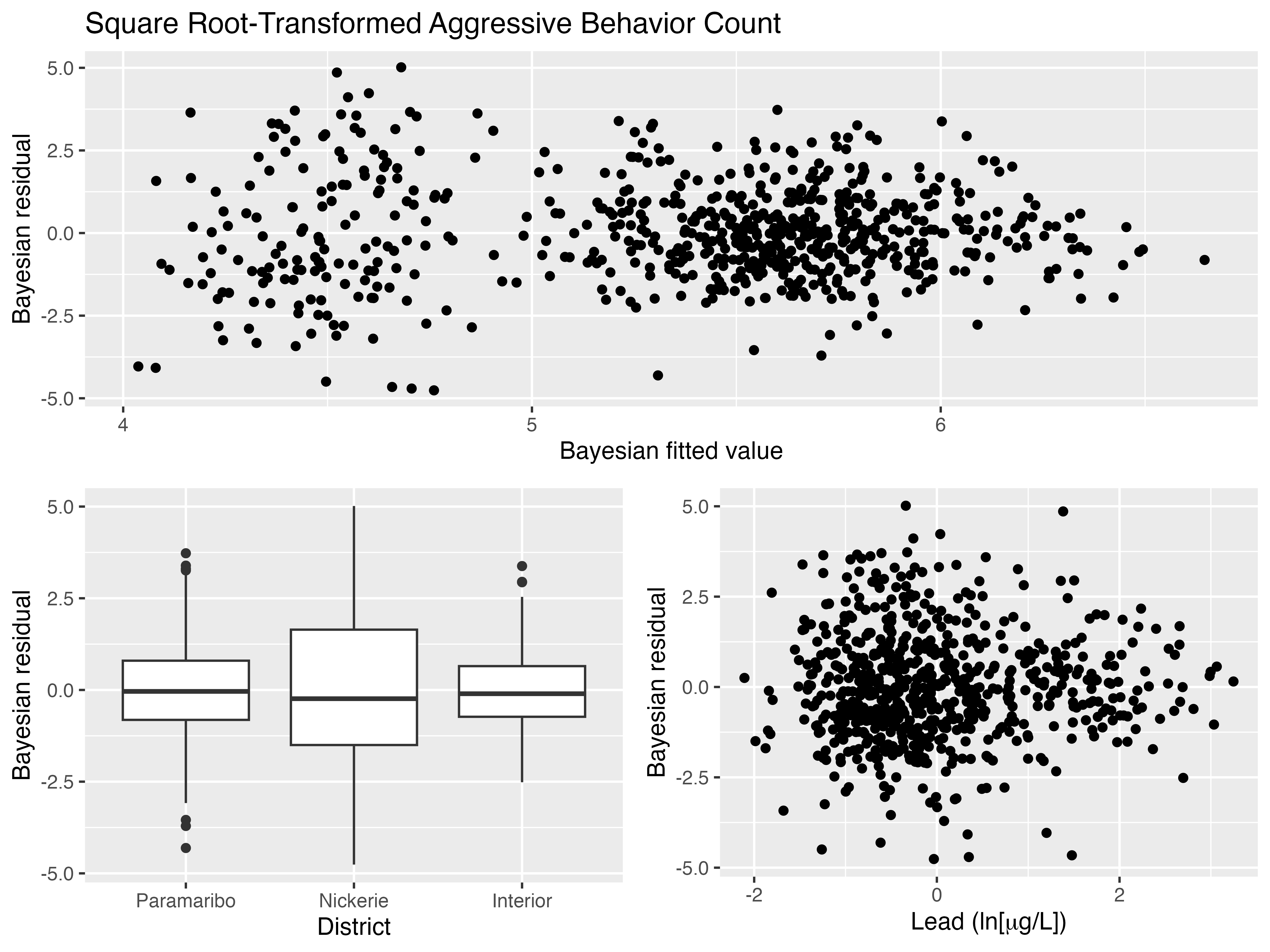}
\caption{BKMR residual diagnostic plots from Case Study 1 for the total problematic behaviors outcome. The top panel displays Bayesian residuals versus Bayesian fitted values. The bottom left panel shows residuals stratified by geographic district, and the bottom right panel displays residuals plotted against transformed Pb concentrations.}
\label{fig:total_diagnostics}
\end{figure}

Figure \ref{fig:total_diagnostics} presents the Bayesian residual patterns from the BKMR model fit. The top panel displays Bayesian residuals plotted against the fitted values. A clear decrease in residual variability is observed with higher dispersion at the lower fitted values. When stratified by district, residual variability is highest in Nickerie, followed by Paramaribo, with the Interior exhibiting the lowest variability. In examining residuals by Pb concentration, a fan-shaped pattern emerges, with variability decreasing as Pb levels increase. There were also possible trends by maternal age and Hg, so these were considered as possible candidate variables in the variance models.

Based on the foregoing assessment of the residual patterns, six HBKMR models were fit, each corresponding to a distinct variance model. These variance models focused on Pb, district, maternal age, and Hg. For Hg, an absolute value transformation was applied due to a diamond-shaped residual pattern. Table \ref{tab:waic_case_study_1_total} presents the WAIC values for the respective models, showing that the HBKMR model with district and Pb included in the $\boldsymbol{W}$ matrix had the lowest WAIC (2494)—a 97-point reduction compared to the BKMR model—indicating a substantial improvement in model fit. Accordingly, subsequent comparisons for this outcome are based on Models 1 (BKMR) and 3 (HBKMR with Pb and district).

\begin{table}[h]
\centering
\caption{Comparison of WAIC values for the models fit in Case Study 1 to the total problematic behaviors and aggressive problematic behaviors outcomes.}
\label{tab:waic_case_study_1_total}
\begin{tabular}{|ll|l|}
\hline
\multicolumn{1}{|l|}{\textbf{Model}} & \textbf{WAIC - Total} & \textbf{WAIC - Aggressive} \\ \hline
\multicolumn{1}{|l|}{1. BKMR} & 2591 & 2018 \\ \hline
\multicolumn{1}{|l|}{2. HBKMR: Pb} & 2585 & 2012 \\ \hline
\multicolumn{1}{|l|}{3. HBKMR: Pb and District} & \textbf{2494} & 2015 \\ \hline
\multicolumn{1}{|l|}{4. HBKMR: Pb and Maternal Age} & 2586 & 1981 \\ \hline
\multicolumn{1}{|l|}{5. HBKMR: Pb, Maternal Age, and District} & 2497 & 1984 \\ \hline
\multicolumn{1}{|l|}{6. HBKMR: \textbar Hg\textbar and District} & 2496 & 1983 \\ \hline
\multicolumn{1}{|l|}{7. HBKMR: District} & 2495 & \textbf{1981} \\ \hline
\multicolumn{1}{|l|}{8. HBKMR: Pb and \textbar Hg\textbar} & 2581 & 2013 \\ \hline
\end{tabular}
\end{table}

\begin{figure}[H]
\centering
\includegraphics[scale = 0.83]{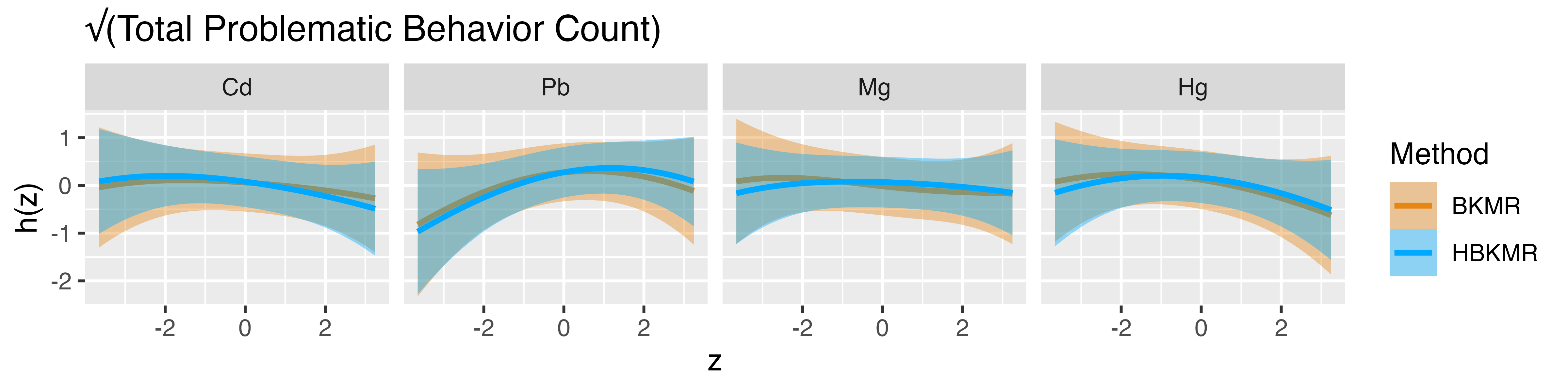}
\caption{Univariate exposure–response function plots comparing the BKMR and HBKMR models in Case Study 1, illustrating their estimated relationships between individual metals and the total problematic behaviors outcome. The plots show point estimates and 95\% credible intervals for each model, with the concentration of each metal varied across its distribution while holding other components at their median levels and adjusting for model covariates.}
\label{fig:univ_case_study_1_total}
\end{figure}

Given the high-dimensional nature of the exposure–response function, visualizations typically rely on selected cross-sections to aid interpretation. Bobb et al\cite{bobb2015bayesian} proposed numerous BKMR cross-sections, including the univariate exposure–response curve, displayed in Figure \ref{fig:univ_case_study_1_total}. This cross-section illustrates the relationship between each individual mixture component and the square-root of total problems. These curves are constructed by varying the concentration of a single metal while holding all others at their median levels, adjusting for the model covariates. The y-axis represents the estimated change in $h(\boldsymbol{z})$, along with the corresponding 95\% credible interval. Across metals, the univariate trends are broadly consistent between BKMR and HBKMR, with HBKMR producing narrower credible intervals and greater precision, particularly at the tails of the exposure distribution.

Among the mixture components, Pb and Hg exhibited the most pronounced associations with the outcome. For Pb, higher concentrations were associated with increased total problems, though the effect plateaued at elevated exposure levels. In contrast, Hg showed a reversed association: lower concentrations were linked to more reported problems, whereas higher levels appeared to protective. This counterintuitive trend may reflect confounding by diet, as Hg exposure in this population likely occurs through fish consumption—a source of both Hg and beneficial nutrients. Thus, higher Hg concentrations may serve as a proxy for better diet quality, potentially mitigating adverse behavioral outcomes. Surprisingly, Cd also exhibited a slightly decreasing relationship, or a ``protective'' effect with respect to total mental health problems. Given the very low prenatal concentrations of Cd in this population, it is important to highlight that these curves reflect very small changes in values and thus these conclusions may not generalize to other populations with higher Cd exposure.  Despite these nuances, both BKMR and HBKMR yield qualitatively similar conclusions regarding individual exposure effects.

\begin{figure}[H]
\centering
\includegraphics[scale = 0.75]{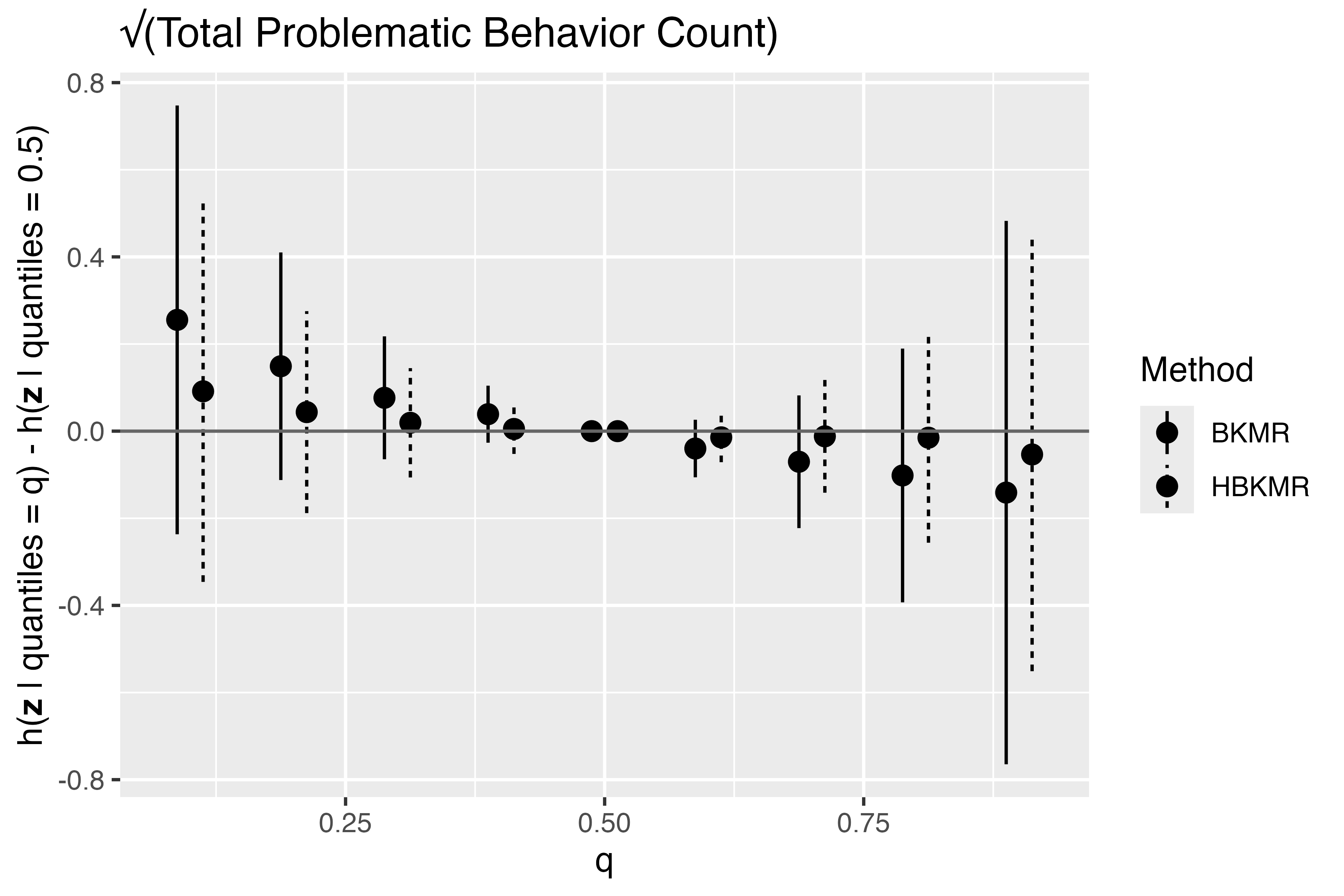}
\caption{Joint effect plots comparing the BKMR and HBKMR models in Case Study 1, illustrating the estimated total problematic behaviors outcome as all mixture components are varied simultaneously from their 50th percentile to their $q*100$th percentile values. The plots show the change in $h(\boldsymbol{z})$ along with the corresponding 95\% credible intervals, adjusting for model covariates.}
\label{fig:overall_case_study_1_total}
\end{figure}

Next, we display the overall mixture effects plot, which captures the joint impact of simultaneously varying all mixture components from their 50th percentile to their $q*100$th percentile values. This quantity, shown in Figure \ref{fig:overall_case_study_1_total}, reflects the estimated change in $h(\boldsymbol{z})$ and its associated 95\% credible interval, adjusted for model covariates. Although the direction and interpretation of the mixture–outcome relationship are consistent across models, HBKMR yielded point estimates that were smaller in magnitude than those from BKMR and produced narrower credible intervals—between 11.1\% and 20.2\% smaller—indicating greater precision. Under both methods, there is little evidence of a statistically meaningful overall mixture effect on total problems, even at the highest exposure contrasts. This apparent null finding may reflect antagonistic effects among mixture components, particularly Pb and Hg, which could offset each other and attenuate the estimated joint effect.

\begin{figure}[H]
\centering
\includegraphics[scale = 0.75]{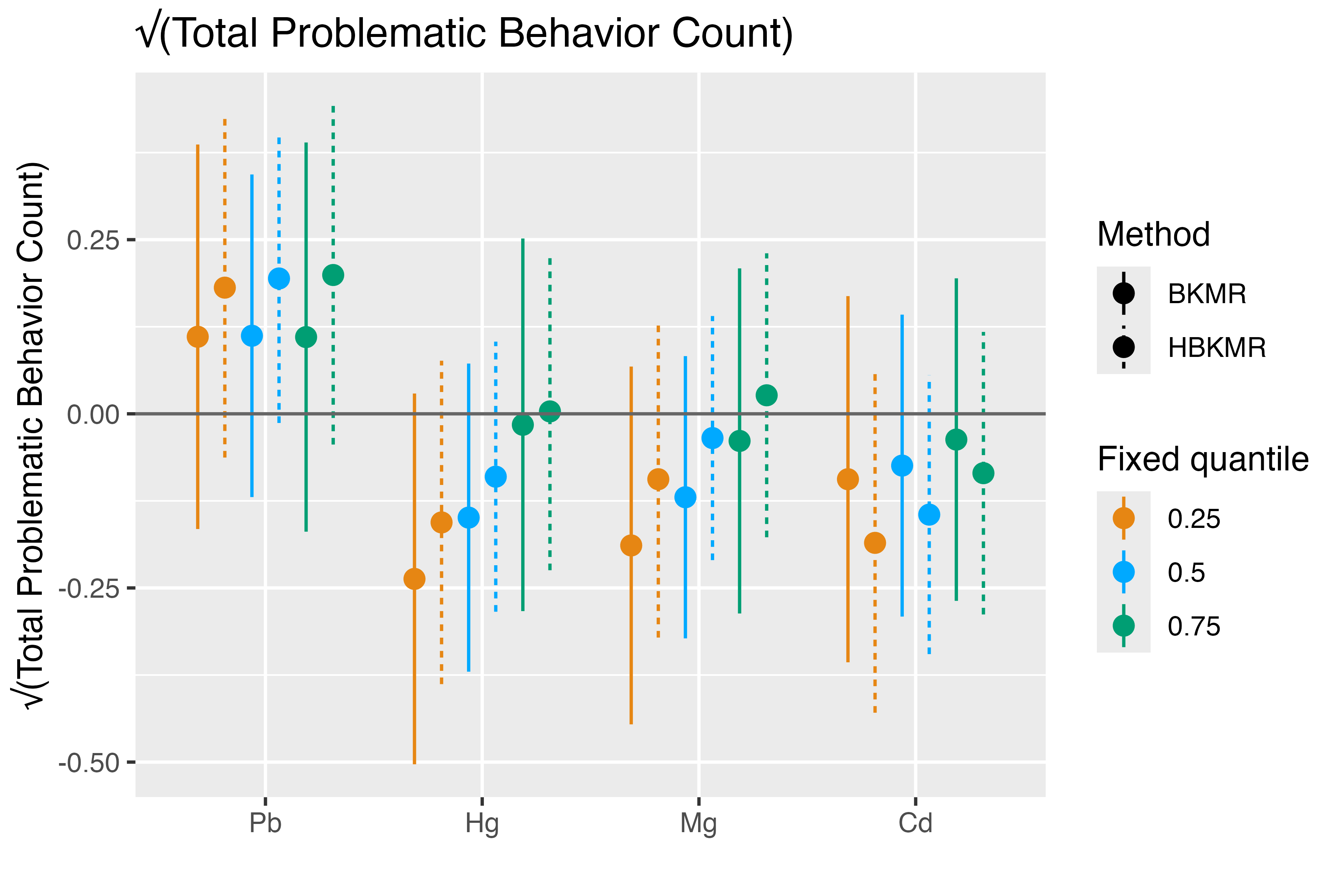}
\caption{Single variable effect plots comparing the BKMR and HBKMR models in Case Study 1, showing the estimated relationship between each individual exposure component and the total problematic behaviors outcome. The plots illustrate the change in $h(\boldsymbol{z})$ and the corresponding 95\% credible intervals as the concentration of each metal is varied from its 25th to 75th percentile, while holding the other components at the color coded ``Fixed quantile" and adjusting for model covariates.}
\label{fig:single_case_study_1_total}
\end{figure}

To elucidate how individual components contribute to the overall mixture effect, Figure \ref{fig:single_case_study_1_total} displays the estimated change in $h(\boldsymbol{z})$ and corresponding 95\% credible interval as each metal is varied from its 25th to 75th percentile, while holding all other components fixed at the levels indicated by the color-coded ``Fixed quantile'' and adjusted for covariates. Point estimates from BKMR and HBKMR generally reflect effects in the same direction, though estimates generated by BKMR and HBKMR differ by a similar factor across metals. Both plots suggest potential interactive effects between metals. Once again, HBKMR yielded narrower credible intervals, offering stronger statistical evidence for a non-negligible effect of Pb that is not as clearly captured by BKMR.

The total problems score provides a global indicator of emotional and behavioral functioning, reflecting the overall burden of difficulties reported by the caregiver. While this composite offers a broad summary, interpretation of general trends benefits from closer examination of key subdomains. Aggressive behavior may be especially relevant in this context, given its salience in early childhood and established links to later behavioral and academic difficulties. The aggressive behaviors domain also exhibited residual patterns suggestive of variance heterogeneity, making it a viable candidate for further analysis using HBKMR. We therefore also examined the association between the metal mixture and the aggressive behavior count in this case study. 

\subsubsection{Aggressive behavioral problems}
An analogous analytical approach was applied to the square-root of the aggressive behaviors count (untransformed median = 10, IQR = 6-15) as was used for square-root of total problems count. The residual diagnostic plots largely mirrored those of the total problems outcome, so the same set of variance specifications was considered for HBKMR. The corresponding WAIC values are shown in Table \ref{tab:waic_case_study_1_total}, which identifies the HBKMR model with district alone in the $\boldsymbol{W}$ matrix as having the best fit, representing a 37-point improvement over the BKMR model. As such, subsequent comparisons for the aggressive behaviors outcome focus on Models 1 and 7.

The univariate exposure–response curves revealed that point estimates from the HBKMR and BKMR models were nearly identical across the range of exposures, with no notable deviations even at the distribution tails. Consistent with the total problems outcome, HBKMR yielded slightly narrower credible intervals compared to BKMR, though the reduction in width was modest. The directional associations between the metals and aggressive behaviors outcome also mirror those of the total count outcome (see the bivariate exposure-response function plots in the Supplementary Materials). Figure \ref{fig:single_case_study_1_aggressive} also shows trends that are largely in line with what was noted for the total problematic behaviors outcome. Here, estimates between HBKMR and BKMR differ, though only slightly, with up to 12\% narrower credible intervals.

\begin{figure}[H]
\centering
\includegraphics[scale = 0.75]{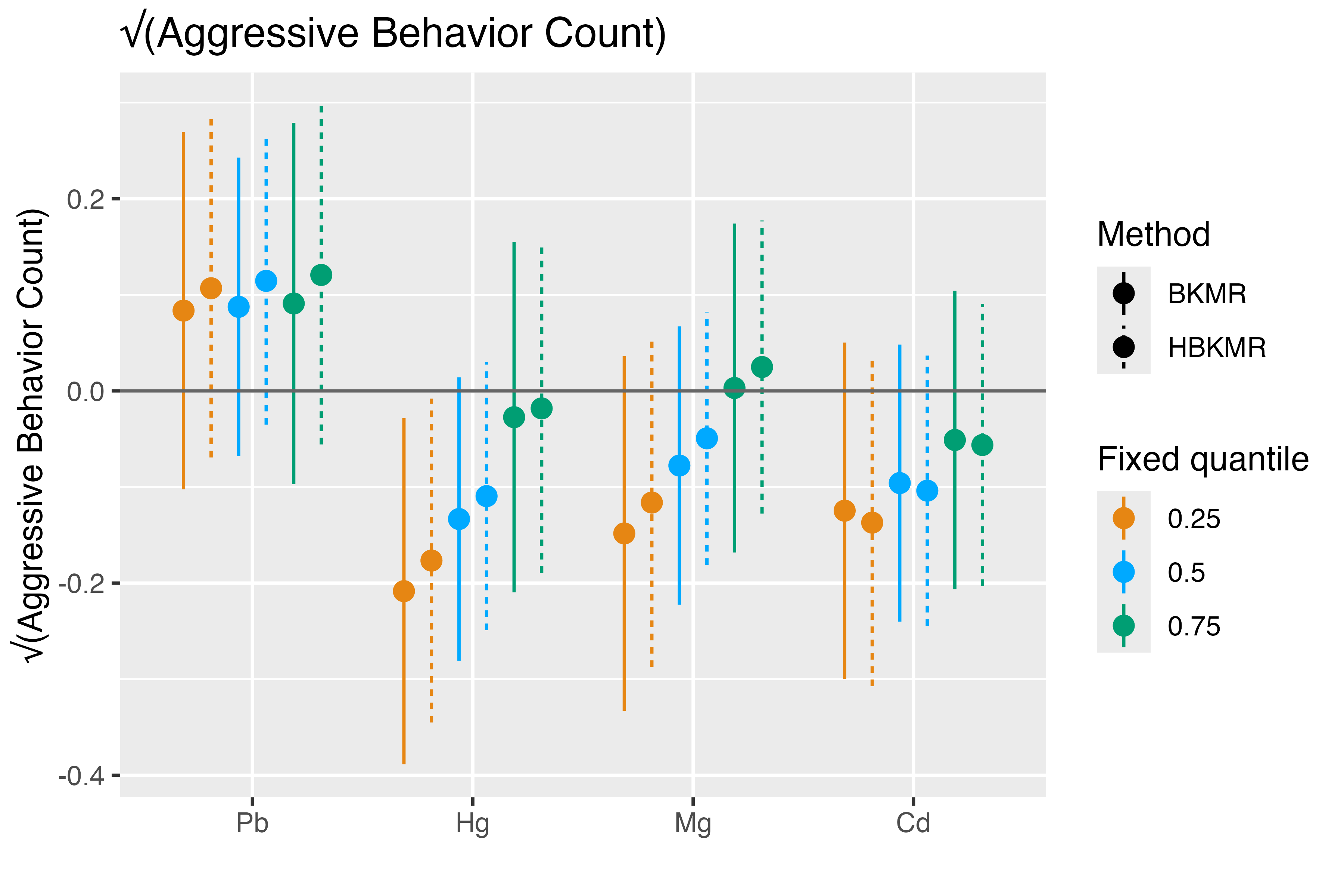}
\caption{Single variable effect plots comparing the BKMR and HBKMR models in Case Study 1, showing the estimated relationship between each individual exposure component and the aggressive behaviors outcome. The plots illustrate the change in $h(\boldsymbol{z})$ and the corresponding 95\% credible intervals as the concentration of each metal is varied from its 25th to 75th percentile, while holding the other components at the color coded ``Fixed quantile" and adjusting for model covariates.}
\label{fig:single_case_study_1_aggressive}
\end{figure}

\begin{figure}[H]
\centering
\includegraphics[scale = 1.6]{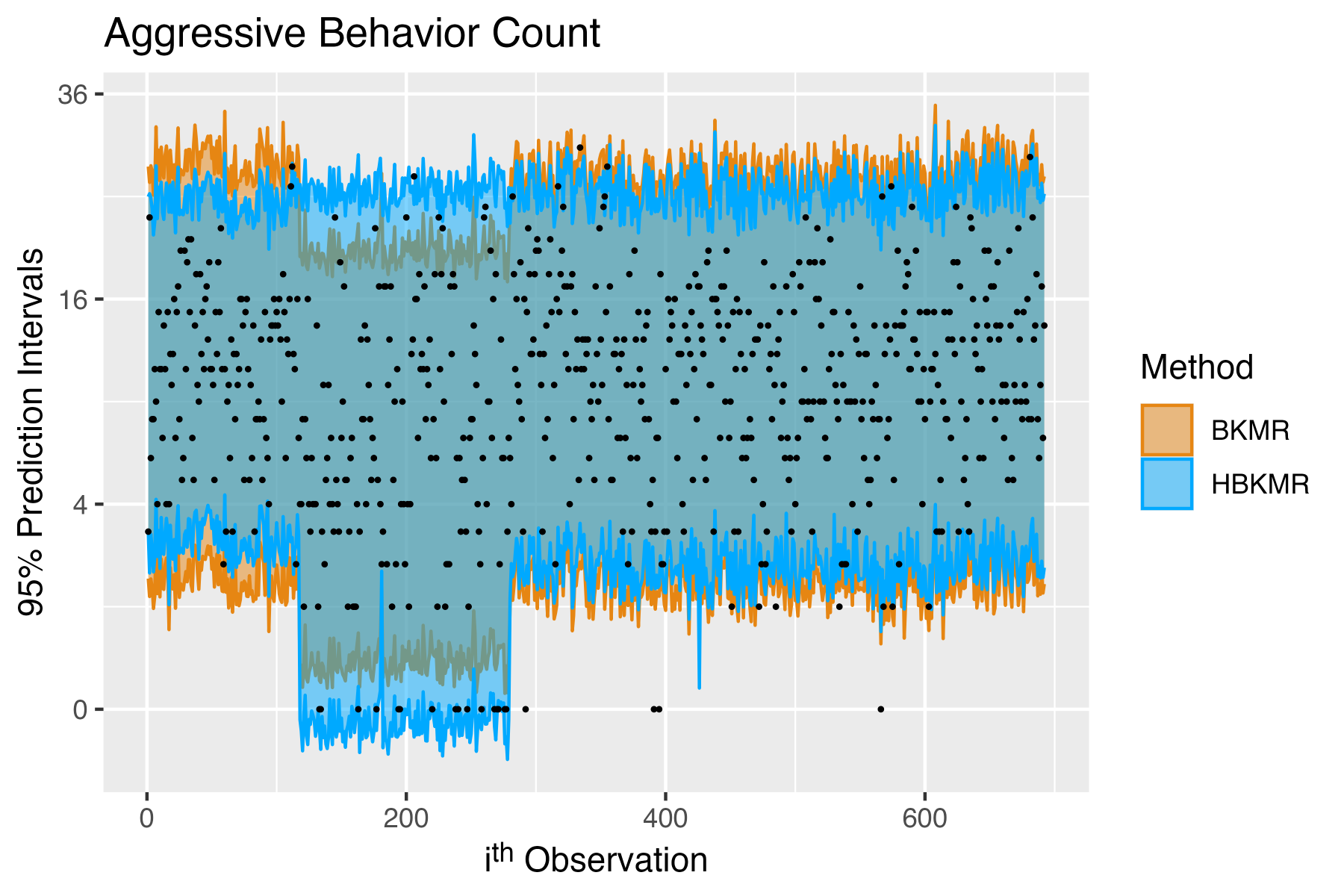}
\caption{95\% posterior predictive intervals comparing the BKMR and HBKMR models in Case Study 1, overlaid with observed values for each observation in the analytical sample.}
\label{fig:prediction_case_study_1_aggressive}
\end{figure}

Lastly, we examined the differences in predictive intervals when using HBKMR and BKMR. Using each model to predict aggressive behavior counts, we obtained 95\% credible intervals for the predictions under each method. As illustrated in Figure \ref{fig:prediction_case_study_1_aggressive}, HBKMR appropriately captured the increased variability associated with outcomes among children from Nickerie, a pattern that the BKMR intervals fail to reflect. For the remaining observations, HBKMR produced narrower predictive intervals compared to BKMR.

\subsection{Metal exposure and reaction time in children living near coal-fired power plants}
We next applied HBKMR to examine the impact of metal exposures on neurobehavioral outcomes in children living near coal-fired power plants in the state of Kentucky. When coal is burned for energy, coal ash, a waste product, is produced. Coal ash is frequently stored in landfills and surface impoundments near marginalized populations.\cite{ref3, ref4} The largest component of coal ash is fly ash, which is mainly comprised of silicon, oxygen, aluminum, iron, and calcium, but also contains trace levels of neurotoxic metals such as arsenic (As), Mn, and Pb.\cite{hatori2010pixe, bednar2013characterization, zierold2020review} Research has shown that fugitive dust emissions from the landfills and surface impoundments can exceed the National Ambient Air Quality Standards for particulate matter $\leq$ 2.5 µm, set by the US Environmental Protection Agency.\cite{epa}  

Fly ash and its neurotoxic components (e.g. Mn) released from the stacks of coal-fired power plants and corresponding storage facilities can impact neurobehavioral development in children.\cite{zierold2022exposure, bjorklund2017manganese, sears2021manganese, lucchini2017manganese, sears2024association} In 2015-2020, research was conducted to investigate the impact of fly ash and metals on neurobehavioral development in children. The methods have been published,\cite{zierold2020protocol} but in brief, fly ash was collected on polycarbonate filters in the homes of children living near two coal-fired power plants in Kentucky for one week. To assess heavy metal body burden, nail samples were collected from 251 children and analyzed by inductively coupled plasma mass spectrometry (ICP-MS). Children’s neurobehavioral performance was assessed by a pediatric psychologist using six tests from the Behavior Assessment Research System (BARS). BARS was developed to provide a series of neurobehavioral tests that are optimized to detect neurotoxicity\cite{rohlman2003development,rohlman2008adaptation} and includes a test measuring simple reaction time (SRT), or a child's motor speed in response to a stimulus on a screen.

In this case study, we sought to determine if a mixture of Mn, Pb, Cd, and As was associated with SRT in these children using BKMR and HBKMR. This analysis included 250 children from this study with complete data on the neurotoxic metal concentrations and SRT. The children were also required to have complete data on gender, age, and maternal education level, as these covariates were relevant to the analysis. One child with an SRT measurement of greater than 1000 miliseconds (ms) was excluded from the analysis due to lack of cooperation on the BARS assessment.

The children in this case study ranged from 6 to 14 years only, and 46\% were female. The majority of children had mothers with at least some college education, while 28\% had mothers who completed a high school degree or less and 13.6\% had mothers who attended graduate school. Toenail metal concentrations were right-skewed with important quantiles displayed in Table \ref{tab:toenail_metals}. Each pair of metals exhibited moderate positive Spearman correlations, with Mn and As experiencing the highest correlation ($r = 0.63$) and Cd and Mn experiencing the lowest correlation ($r = 0.41$). Variance inflation factors suggested that there was no evidence of multicollinearity in this analysis.

\begin{table}[h]
 \centering
\caption{Quantiles of toenail metal concentrations in the 250 children living near coal-fired power plants.}
\label{tab:toenail_metals}
\begin{tabular}{|l|l|l|l|l|l|}
\hline
\textbf{\begin{tabular}[c]{@{}l@{}}Toenail metal \\ concentration \\ (ppm)\end{tabular}} & \textbf{10\%} & \textbf{25\%} & \textbf{50\%} & \textbf{75\%} & \textbf{90\%} \\ \hline
As                                                                                       & 0.05          & 0.07          & 0.11          & 0.22          & 0.35          \\ \hline
Cd                                                                                       & 0.01          & 0.02          & 0.04          & 0.08          & 0.17          \\ \hline
Mn                                                                                       & 0.37          & 0.59          & 0.99          & 1.78          & 2.94          \\ \hline
Pb                                                                                       & 0.16          & 0.28          & 0.47          & 0.87          & 1.74          \\ \hline
\end{tabular}
\end{table}

For stability, SRT was converted to seconds prior to fitting the models, but all mixture effects were computed in milliseconds. Models were adjusted for child's gender, child's age in years (treated continuously), and mother's education level.
\begin{figure}[H]
\centering 
\includegraphics[scale = 0.63]{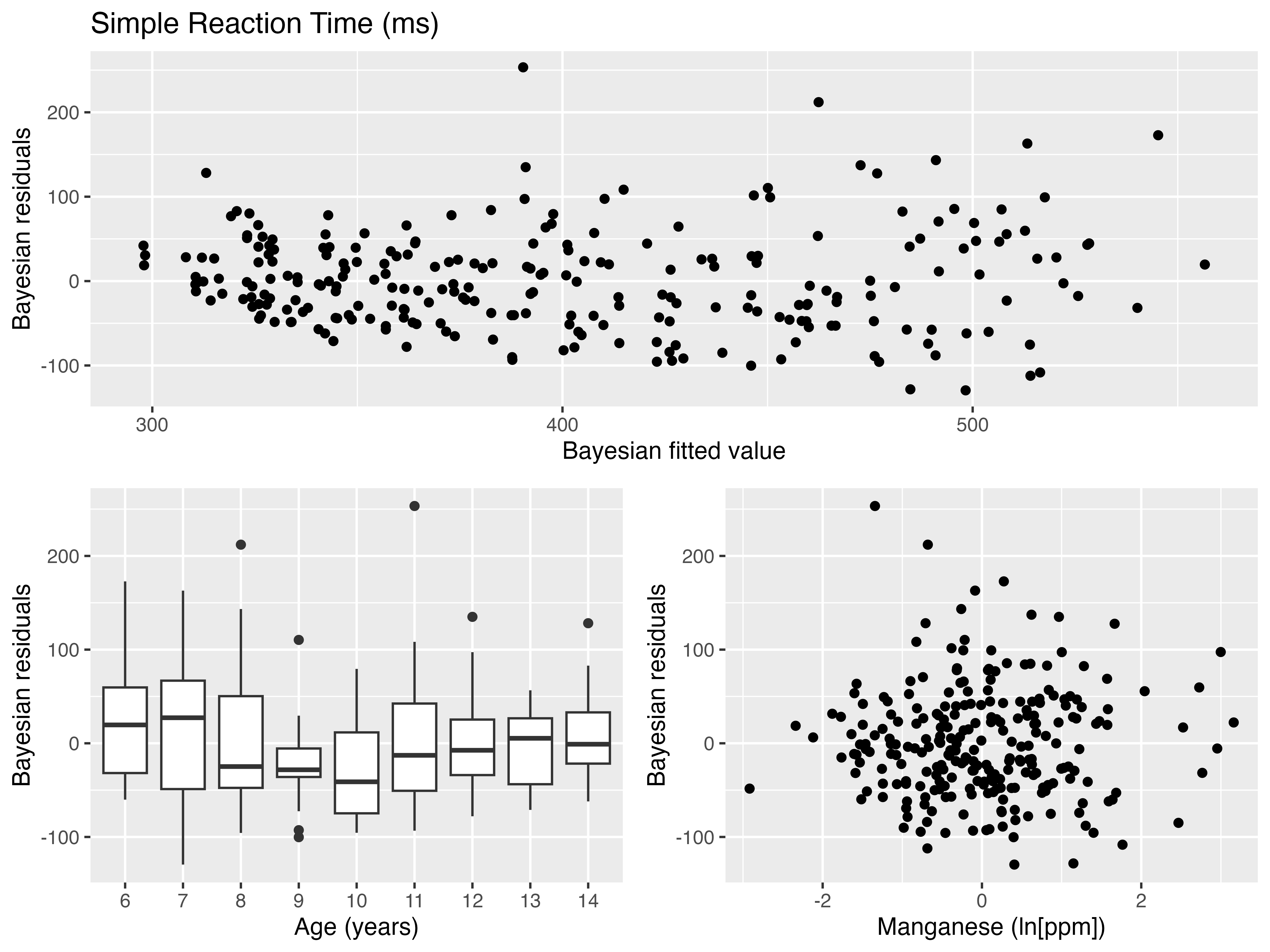}
\caption{BKMR residual diagnostic plots from Case Study 2. The top plot displays Bayesian residuals versus Bayesian fitted values from the BKMR model fit. The bottom left plot displays the Bayesian residuals versus age of the children in years, and the bottom right plot displays the Bayesian residuals versus transformed toenail Mn concentrations.}
\label{fig:srt_diagnostics}
\end{figure}

Figure \ref{fig:srt_diagnostics} illustrates the Bayesian residual diagnostic plots from the BKMR model fit (Model 1). The top panel in this figure depicts the Bayesian residuals vs. fitted values. A fan shape is evident, with an increasing variance as the fitted values increase for SRT. Furthermore, Figure \ref{fig:srt_diagnostics} includes two plots displaying Bayesian residuals vs. model predictors. The bottom left plot illustrates a non-constant variance in the residuals by the age of the child in years, with more variability typically observed at lower ages. The bottom right plot displays the Bayesian residuals vs. transformed Mn values, with a possible increase in variance of the residuals at higher Mn levels. With respect to the remaining metals, no large differences in residual variances were observed.

After examining the diagnostic plots, we fit two different HBKMR models, considering age and Mn as candidates in the HBKMR variance model. In the first HBKMR model (Model 2), the $\boldsymbol{W}$ matrix included child's age (treated continuously), and in the second model, the $\boldsymbol{W}$ matrix included child's age and transformed Mn concentration. Table \ref{tab:waic_case_study_2} displays the WAIC value for each of these model fits, suggesting that Model 2, considering child's age in the variance model, fit best. Notably, this HBKMR model had a WAIC of -677, nearly 30 points lower than the WAIC of the BKMR model. Thus, comparisons were made between Models 1 and 2 in the remainder of the case study.

\begin{figure}[H]
\centering
\includegraphics[scale = 0.85]{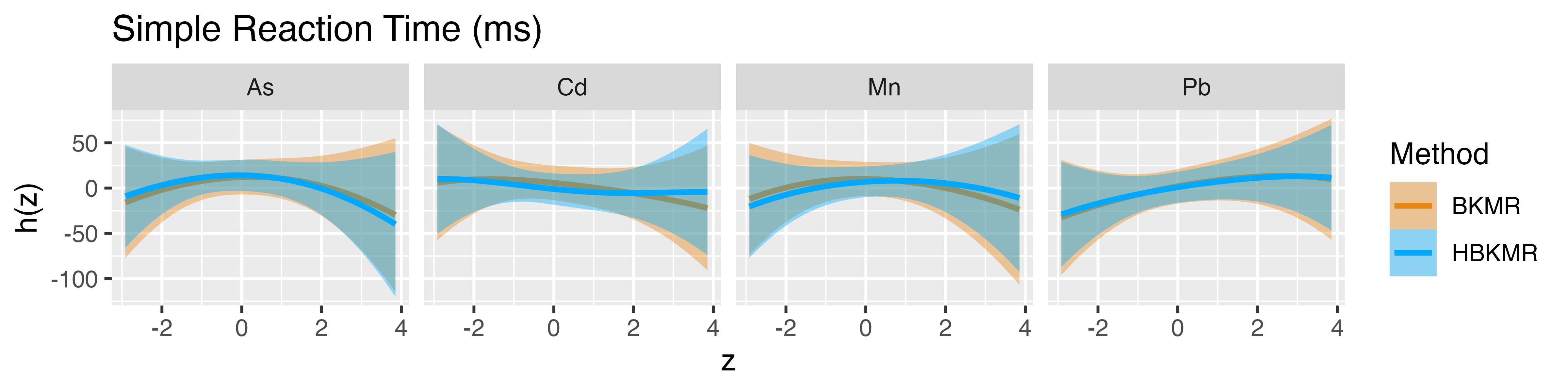}
\caption{Univariate exposure-response function plots comparing BKMR and HBKMR in Case Study 2. Plots depict the change in estimated $h(\boldsymbol{z})$, and the corresponding 95\% credible interval, when varying the value of the specified metal concentration while fixing the remaining metals at their medians and controlling for child's age, child's gender, and maternal education level.}
\label{fig:univ_case_study}
\end{figure}

\begin{figure}[H]
\centering
\includegraphics[scale = 0.9]{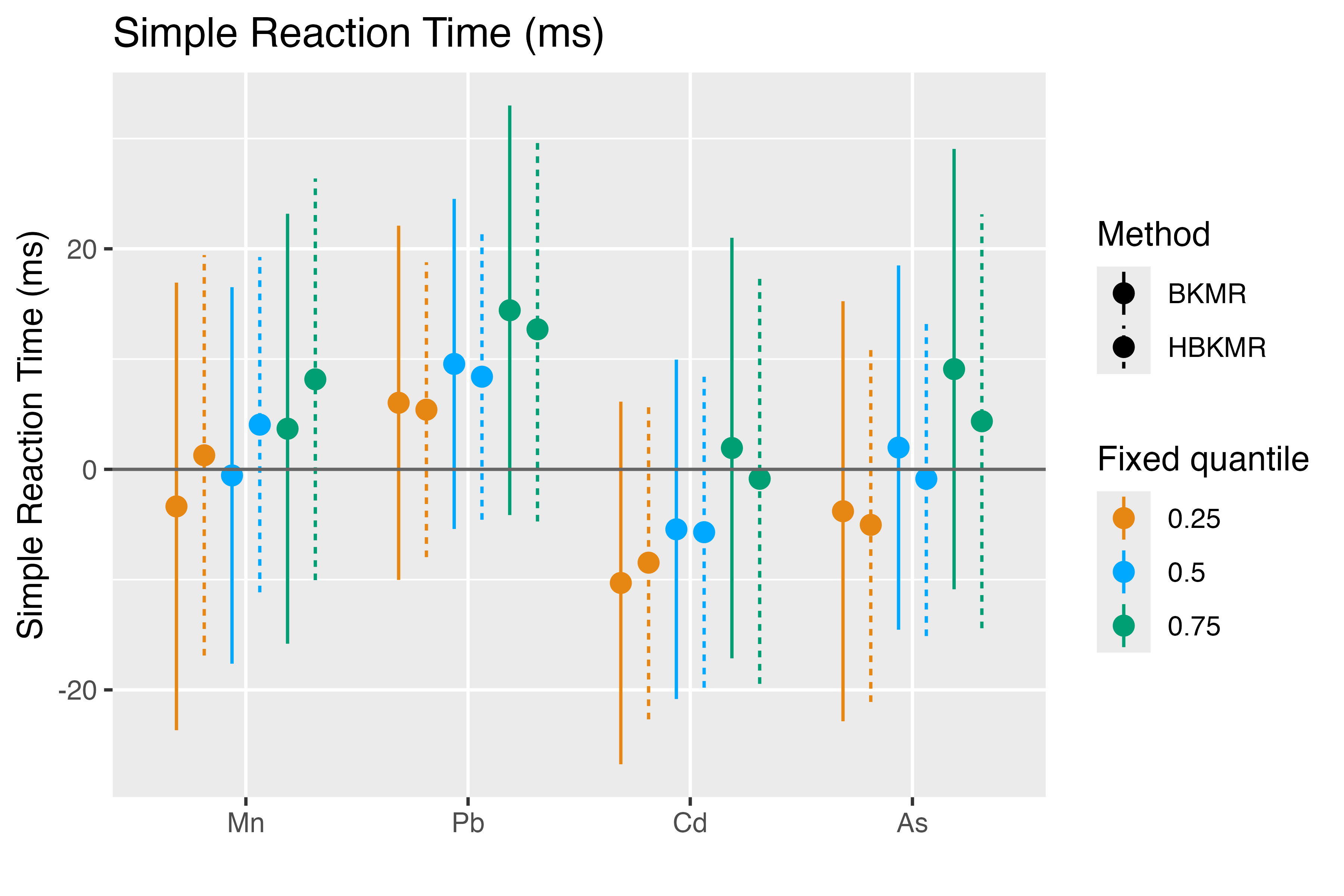}
\caption{Single variable effect plots comparing BKMR and HBKMR in Case Study 2. Plots depict the change in estimated $h(\boldsymbol{z})$, and the corresponding 95\% credible interval, when changing the metal of interest from its 25th percentile to its 75th percentile while holding the remaining metals fixed at the color-coded ``Fixed quantile.'' Estimates are adjusted for child’s age, child’s gender, and maternal education level.}
\label{fig:single_variable_case_study_2}
\end{figure}

\begin{figure}[H]
\centering
\includegraphics[scale = 0.9]{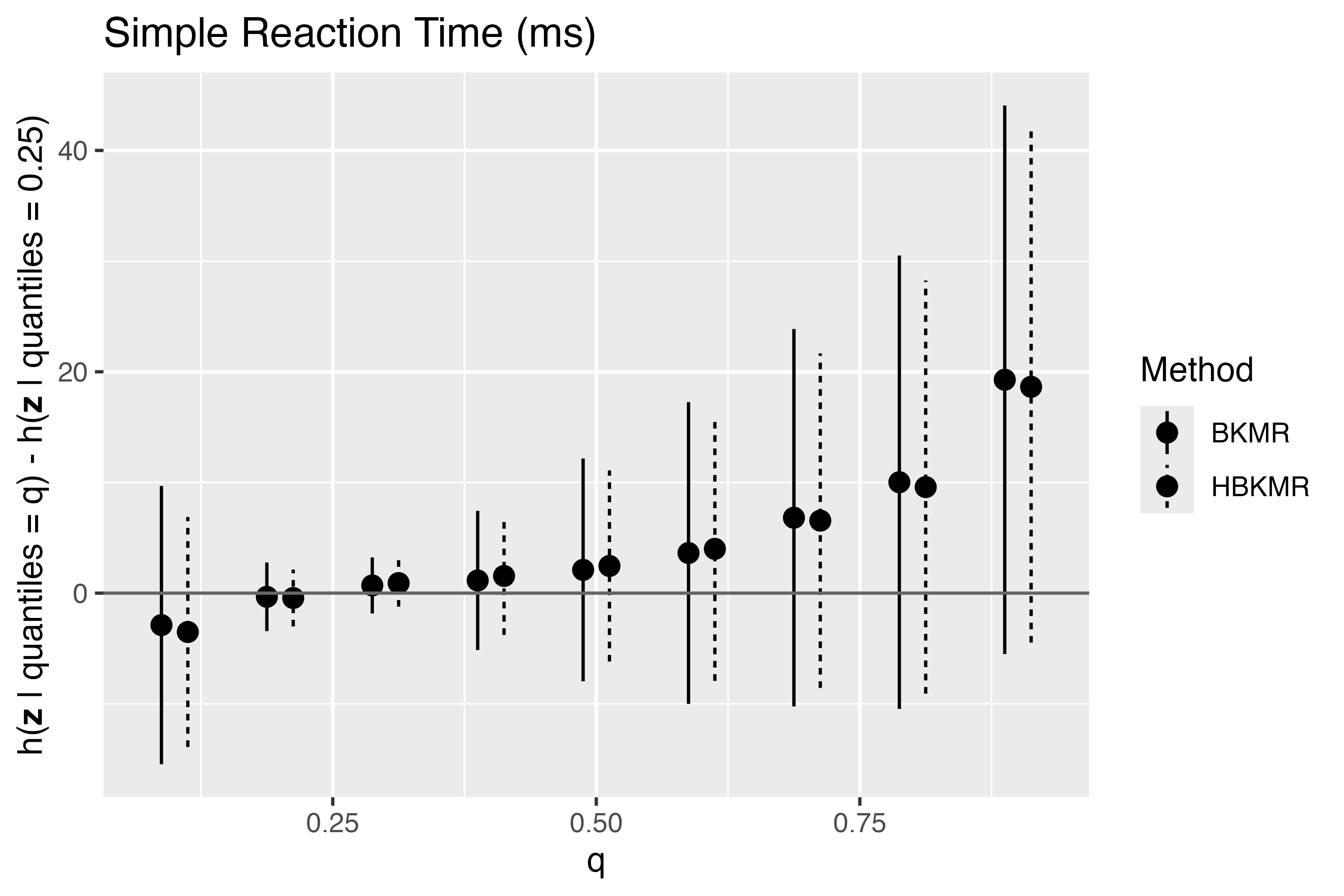}
\caption{Joint effect plots comparing BKMR and HBKMR in Case Study 2. Plots depict the change in $h(\boldsymbol{z})$ when simultaneously changing all of the metals from their 25th percentile values to their $q*100$th percentile values and holding child's age, child's gender, and maternal education level fixed. Posterior means are plotted along with 95\% credible intervals.}
\label{fig:overall_case_study_2}
\end{figure}

\begin{table}[h]
\centering
\caption{Comparison of WAIC values for the models fit in Case Study 2.}
\label{tab:waic_case_study_2}
\begin{tabular}{|ll|}
\hline
\multicolumn{1}{|l|}{\textbf{Model}}       & \textbf{WAIC} \\ \hline
\multicolumn{1}{|l|}{1. BKMR}              & -648          \\ \hline
\multicolumn{1}{|l|}{2. HBKMR: Age}        & -677          \\ \hline
\multicolumn{1}{|l|}{3. HBKMR: Age and Mn} & -676          \\ \hline
\end{tabular}
\end{table}
The univariate exposure-response function plots with 95\% credible intervals are displayed in Figure \ref{fig:univ_case_study}. Under both methods, these curves suggest that there appear to be only modest relationships between each of the metals and SRT after adjusting for the covariates and holding the remaining metals constant at their median values. Increasing levels of Pb may be associated with higher SRT, or slower reaction times. Slight upside-down U shapes may also be possible, especially with As. The general patterns typically aligned between BKMR and HBKMR, though point estimates of exposure-response cross-sections differed in some cases. For example, in Figure \ref{fig:univ_case_study} the point estimates for the exposure-response function when changing Cd but holding all of the remaining metals at their medians become flatter under HBKMR, with the values of these curves deviating from BKMR in the regions with greater uncertainty. The 95\% credible intervals for these curves are almost always narrower under HBKMR, albeit by a small amount, with the most notable differences towards the extremes of the metal distributions. The single variable effect plots shown in Figure \ref{fig:single_variable_case_study_2} also illustrated differing point estimates than BKMR, with estimates often falling closer to the null and narrower 95\% credible intervals.
Figure \ref{fig:overall_case_study_2} displays a joint effect plot comparing HBKMR and BKMR. These plots depict the change in $h(\boldsymbol{z})$ when changing all of the metal concentration values simultaneously from their 25th percentile values to their $q*100$th percentile values, while holding the sociodemographic variables constant. Notably, while the point estimates are very similar between BKMR and HBKMR, the 95\% credible interval widths are narrower under HBKMR, leading to more precise inferences on the joint effects. There is approximately a 6 - 17\% decrease in credible interval width when using HBKMR  vs. BKMR, depending on the joint effect that is plotted, with the largest differences observed at the lower quantiles. These plots highlight the advantage of using HBKMR, which required a variance model with only one covariate (child's age) to better account for heteroscedasticity in the error terms. While overall conclusions generally remained the same under BKMR and HBKMR in this data example, a 17\% reduction in credible interval width in a dataset with more signal could affect another study's conclusions.

This joint effect plot suggests that there is plausible evidence of a non-zero effect of the metal mixture on SRT when changing all of the metal concentrations from their 25th percentile to 90th percentile values under HBKMR and BKMR. This effect is estimated to be approximately 20 ms with a 95\% credible interval that just crosses 0 under both BKMR and HBKMR. However, when examined in combination with the single variable plot, this joint effect appears to be almost entirely described by Pb. There is substantial uncertainty across many of the effects in these plots, though HBKMR tends to produce narrower credible intervals than BKMR.

In this analysis, we utilized an Inverse-Uniform(0, 100) prior on each $r$ parameter. This was the most informative prior used in this case study, so we sought to determine how comparisons between BKMR and HBKMR might change under a more informative prior distribution. We considered a weakly-informative Uniform(0,5) prior on each $r$ parameter and compared the point estimates and 95\% credible interval widths for the joint effects in a sensitivity analysis. There remained strong similarities in the joint effect estimates between BKMR and HBKMR under this less informative prior distribution, and the relative change in credible interval width also remained similar (between 5-15\%). Point estimates of the joint effects were similar, though slightly lower, when using the less informative Uniform(0,5) prior compared to the default Inverse-Uniform(0,100) prior. See Table \ref{tab:sensitivity_analysis} for a full comparison of point estimates and 95\% credible intervals in this sensitivity analysis.

\begin{sidewaystable}[h]
\caption{Sensitivity analysis displaying the joint effect estimates, 95\% credible interval widths, and reduction in 95\% credible interval width using HBKMR vs. BKMR in Case Study 2. The influence of an informative inverse-uniform prior for $\boldsymbol{r}$ and a weakly-informative uniform prior for $\boldsymbol{r}$ are considered.}
\label{tab:sensitivity_analysis}
\resizebox{\textwidth}{!}{%
\begin{tabular}{|l|lllll|lllll|}
\hline
\multicolumn{1}{|c|}{\multirow{2}{*}{Sensitivity analysis}} & \multicolumn{5}{c|}{\textbf{$r \sim$ Inverse-Uniform(0,100)}}                                                                                                                                                                                                                                 & \multicolumn{5}{c|}{\textbf{$r\sim$ Uniform(0,5)}}                                                                                                                                                                                                                                            \\ \cline{2-11} 
\multicolumn{1}{|c|}{}                                      & \multicolumn{2}{c|}{\textbf{BKMR}}                                                   & \multicolumn{3}{c|}{\textbf{HBKMR}}                                                                                                                                                                    & \multicolumn{2}{c|}{\textbf{BKMR}}                                                   & \multicolumn{3}{c|}{\textbf{HBKMR}}                                                                                                                                                                    \\ \hline
\multicolumn{1}{|c|}{\textbf{Joint Effect}}                 & \multicolumn{1}{c|}{\textbf{Estimate}} & \multicolumn{1}{c|}{\textbf{95\% CI width}} & \multicolumn{1}{c|}{\textbf{Estimate}} & \multicolumn{1}{c|}{\textbf{95\% CI width}} & \multicolumn{1}{c|}{\textbf{\begin{tabular}[c]{@{}c@{}}\% decrease in \\ CI width using \\ HBKMR\end{tabular}}} & \multicolumn{1}{c|}{\textbf{Estimate}} & \multicolumn{1}{c|}{\textbf{95\% CI width}} & \multicolumn{1}{c|}{\textbf{Estimate}} & \multicolumn{1}{c|}{\textbf{95\% CI width}} & \multicolumn{1}{c|}{\textbf{\begin{tabular}[c]{@{}c@{}}\% decrease in \\ CI width using \\ HBKMR\end{tabular}}} \\ \hline
$h(\boldsymbol{z}_{0.10}) - h(\boldsymbol{z}_{0.25})$       & \multicolumn{1}{l|}{-2.89}             & \multicolumn{1}{l|}{2.51}                   & \multicolumn{1}{l|}{-3.51}             & \multicolumn{1}{l|}{2.08}                   & 17.4                                                                                                            & \multicolumn{1}{l|}{-4.23}             & \multicolumn{1}{l|}{2.13}                   & \multicolumn{1}{l|}{-4.53}             & \multicolumn{1}{l|}{1.80}                   & 15.7                                                                                                            \\ \hline
$h(\boldsymbol{z}_{0.20}) - h(\boldsymbol{z}_{0.25})$       & \multicolumn{1}{l|}{-0.33}             & \multicolumn{1}{l|}{0.62}                   & \multicolumn{1}{l|}{-0.44}             & \multicolumn{1}{l|}{0.51}                   & 17.1                                                                                                            & \multicolumn{1}{l|}{-0.78}             & \multicolumn{1}{l|}{0.53}                   & \multicolumn{1}{l|}{-0.80}             & \multicolumn{1}{l|}{0.46}                   & 14.4                                                                                                            \\ \hline
$h(\boldsymbol{z}_{0.30}) - h(\boldsymbol{z}_{0.25})$       & \multicolumn{1}{l|}{0.69}              & \multicolumn{1}{l|}{0.51}                   & \multicolumn{1}{l|}{0.91}              & \multicolumn{1}{l|}{0.43}                   & 16.0                                                                                                            & \multicolumn{1}{l|}{1.11}              & \multicolumn{1}{l|}{0.44}                   & \multicolumn{1}{l|}{1.29}              & \multicolumn{1}{l|}{0.38}                   & 13.2                                                                                                            \\ \hline
$h(\boldsymbol{z}_{0.40}) - h(\boldsymbol{z}_{0.25})$       & \multicolumn{1}{l|}{1.14}              & \multicolumn{1}{l|}{1.26}                   & \multicolumn{1}{l|}{1.55}              & \multicolumn{1}{l|}{1.07}                   & 15.3                                                                                                            & \multicolumn{1}{l|}{2.27}              & \multicolumn{1}{l|}{1.09}                   & \multicolumn{1}{l|}{2.59}              & \multicolumn{1}{l|}{0.96}                   & 11.5                                                                                                            \\ \hline
$h(\boldsymbol{z}_{0.50}) - h(\boldsymbol{z}_{0.25})$       & \multicolumn{1}{l|}{2.10}              & \multicolumn{1}{l|}{2.01}                   & \multicolumn{1}{l|}{2.45}              & \multicolumn{1}{l|}{1.73}                   & 14.2                                                                                                            & \multicolumn{1}{l|}{3.75}              & \multicolumn{1}{l|}{1.74}                   & \multicolumn{1}{l|}{4.05}              & \multicolumn{1}{l|}{1.56}                   & 10.2                                                                                                            \\ \hline
$h(\boldsymbol{z}_{0.60}) - h(\boldsymbol{z}_{0.25})$       & \multicolumn{1}{l|}{3.62}              & \multicolumn{1}{l|}{2.73}                   & \multicolumn{1}{l|}{4.00}              & \multicolumn{1}{l|}{2.38}                   & 12.5                                                                                                            & \multicolumn{1}{l|}{5.52}              & \multicolumn{1}{l|}{2.34}                   & \multicolumn{1}{l|}{5.94}              & \multicolumn{1}{l|}{2.14}                   & 8.4                                                                                                             \\ \hline
$h(\boldsymbol{z}_{0.70}) - h(\boldsymbol{z}_{0.25})$       & \multicolumn{1}{l|}{6.81}              & \multicolumn{1}{l|}{3.41}                   & \multicolumn{1}{l|}{6.55}              & \multicolumn{1}{l|}{3.02}                   & 11.4                                                                                                            & \multicolumn{1}{l|}{8.27}              & \multicolumn{1}{l|}{2.95}                   & \multicolumn{1}{l|}{8.24}              & \multicolumn{1}{l|}{2.75}                   & 7.0                                                                                                             \\ \hline
$h(\boldsymbol{z}_{0.80}) - h(\boldsymbol{z}_{0.25})$       & \multicolumn{1}{l|}{10.02}             & \multicolumn{1}{l|}{4.10}                   & \multicolumn{1}{l|}{9.58}              & \multicolumn{1}{l|}{3.73}                   & 8.9                                                                                                             & \multicolumn{1}{l|}{10.39}             & \multicolumn{1}{l|}{3.60}                   & \multicolumn{1}{l|}{10.56}             & \multicolumn{1}{l|}{3.40}                   & 5.5                                                                                                             \\ \hline
$h(\boldsymbol{z}_{0.90}) - h(\boldsymbol{z}_{0.25})$       & \multicolumn{1}{l|}{19.28}             & \multicolumn{1}{l|}{4.96}                   & \multicolumn{1}{l|}{18.64}             & \multicolumn{1}{l|}{4.62}                   & 6.7                                                                                                             & \multicolumn{1}{l|}{16.31}             & \multicolumn{1}{l|}{4.53}                   & \multicolumn{1}{l|}{16.31}             & \multicolumn{1}{l|}{4.27}                   & 5.8                                                                                                             \\ \hline
\end{tabular}%
}
\end{sidewaystable}

\section{Discussion}  \label{sec:discussion}
In this paper, we introduced HBKMR and compared it to BKMR through two real-world case studies involving heteroscedastic health outcomes. In both case studies, HBKMR exhibited a notable improvement in fit, with differences in WAIC values ranging from 29 to 97 compared to BKMR. Through newly-created BKMR diagnostic plotting functions, we identified model covariates that explained differences in the variance of the residuals and incorporated these covariates into an error term variance model through the HBKMR $\boldsymbol{W}$ matrix. Importantly, utilizing a variance model led to more precise inferences around cross-sections of the $h(\cdot)$ function and around calculated health effects of the environmental mixture. It also led to differences in point estimates for certain exposure-response cross sections, particularly when considering the square-root of the total behavioral outcomes in Case Study 1 and simple reaction time in Case Study 2. 

Inferences around posterior predictions also differed under BKMR and HBKMR. Children with lower levels of prenatal Pb exposure and children living within the Nickerie district in Case Study 1 had more residual variability than other children in the study, and this was appropriately captured in the 95\% posterior prediction intervals for the square-root of total mental health problems and the square-root of aggressive behavioral problems. While BKMR is most often used to quantify inferences on the individual and mixture effects of exposures (i.e. the $h(\cdot)$ function), it has also been used to predict counterfactual health outcomes to estimate causal effects.\cite{devick2022bayesian} The use of HBKMR in such settings, when outcomes exhibit heteroscedasticity, could improve uncertainty estimates around causal effects. There is also promise in using HBKMR for prediction in the context of precision environmental health. When heteroscedasticity is present, HBKMR could enable researchers to appropriately estimate predicted health outcomes for individuals and quantify uncertainty in these predictions, considering an individual's unique characteristics and exposure profiles.

HBKMR and the present study have several important strengths. Our new diagnostic plotting functions allow researchers to check the model assumptions for BKMR, a step that is often overlooked in practice. These plots enable researchers to check the homoscedasticity assumption and identify covariates that can be used in a variance model, if the homoscedasticy assumption is violated. Importantly, this function could also be used to examine the appropriateness of the Gaussian kernel function, a choice that is often accepted but not scrutinized. If non-linear relationships between Bayesian residuals and exposure values in the mixture are observed, this could suggest that an alternative kernel function may be preferred. As noted, we demonstrated improvements in model fit, tighter 95\% credible intervals, and differences in point estimates in two real-world applications with heteroscedasticity. Our flexible implementation of both HBKMR and BKMR in the NIMBLE is another noteworthy strength of this work and allows for an apples-to-apples comparison between HBKMR and BKMR as well as more flexibility in the specified prior distributions. We observed an improvement in computational efficiency and convergence to the posterior in our applications when using our NIMBLE BKMR implementation, without the need to tune Metropolis-Hastings steps. Tuning the Metropolis-Hastings steps is often necessary when using the BKMR R package to ensure reasonable acceptance rates for the $\boldsymbol{r}$ parameters. \cite{bobb2018statistical} 

In addition to our study's strengths, there are also important limitations to discuss. First, we observed differences in point estimates with certain outcomes but not others, and an in-depth simulation study could uncover settings in which differences in point estimates may be most pronounced. We also examined a limited number of prior distributions on the $\boldsymbol{r}$ parameters, and Bobb et al noted that while relative variable importance often stays the same with choice of prior, the magnitudes of computed effects could differ.\cite{bobb2018statistical} This limitation extends to HBKMR as well. We did, however, apply both a quite informative and quite non-informative prior to $\boldsymbol{r}$, which yielded similar conclusions about the joint effects and the relative appropriateness of each model when comparing HBKMR and BKMR in Case Study 2. Overall, this work illustrates the utility of HBKMR and the importance of checking the BKMR's model assumptions through two in-depth case studies involving metal exposures, children's neurodevelopment, and children's neurobehavioral function.

\section*{Acknowledgements}
This work was supported by Grant No. UL1TR003096, NCATS through the CCTS Statistical and Analytic Methods Development Pilot Grant Program, Grant No. U01TW010087 and U2RTW010104, Fogarty International Center of NIH, and Grant No. R01ES024757, NIEHS.

\section*{Data Availability Statement}
To preserve data confidentiality, we are not able to share the datasets in the case studies. However, the R code to diagnose heteroscedasticity and implement HBKMR on a simulated dataset is available in the following repository: \url{https://github.com/iebuker/hbkmr_example}. Furthermore, an HBKMR tutorial is available at the following link: \url{https://iebuker.github.io/hbkmr_example/}.

\bibliographystyle{vancouver}
\bibliography{references}

\section*{Appendix}
Here, we provide the derivation of the posterior distribution of $\boldsymbol{\beta}, \tau, \boldsymbol{r},$ and $\boldsymbol{\gamma}$. This posterior distribution results from integrating $\boldsymbol{h}$ out of the full posterior distribution. Recall that the full posterior distribution, including $\boldsymbol{h}$, can be specified as:
\begin{align*}
p(\boldsymbol{\beta}, \tau, \boldsymbol{r}, \boldsymbol{h}, \boldsymbol{\gamma}\mid \boldsymbol{y}, \boldsymbol{X}, \boldsymbol{Z}, \boldsymbol{W}) 
&\propto p(\boldsymbol{y} \mid \boldsymbol{\beta}, \boldsymbol{h}, \boldsymbol{\gamma}, \boldsymbol{X}, \boldsymbol{W})p(\boldsymbol{h}\mid \tau, \boldsymbol{r}, \boldsymbol{Z})p(\boldsymbol{\beta})p(\boldsymbol{r})p(\tau)p(\boldsymbol{\gamma})
\end{align*}
Thus,
\begin{align*}
p(\boldsymbol{\beta}, \tau, \boldsymbol{r}, \boldsymbol{\gamma}\mid \boldsymbol{y}, \boldsymbol{X}, \boldsymbol{Z}, \boldsymbol{W}) 
&\propto \int p(\boldsymbol{y} \mid \boldsymbol{\beta}, \boldsymbol{h}, \boldsymbol{\gamma}, \boldsymbol{X}, \boldsymbol{W})p(\boldsymbol{h}\mid \tau, \boldsymbol{r}, \boldsymbol{Z})p(\boldsymbol{\beta})p(\boldsymbol{r})p(\tau)p(\boldsymbol{\gamma}) d\boldsymbol{h} \\
&\propto p(\boldsymbol{\beta})p(\boldsymbol{r})p(\tau)p(\boldsymbol{\gamma}) \int p(\boldsymbol{y} \mid \boldsymbol{\beta}, \boldsymbol{h}, \boldsymbol{\gamma}, \boldsymbol{X}, \boldsymbol{W})p(\boldsymbol{h}\mid \tau, \boldsymbol{r}, \boldsymbol{Z}) d\boldsymbol{h}
\end{align*}
Note that the kernel of $p(\boldsymbol{y} \mid \boldsymbol{\beta}, \boldsymbol{h}, \boldsymbol{\gamma}, \boldsymbol{X}, \boldsymbol{W})p(\boldsymbol{h}\mid \tau, \boldsymbol{r}, \boldsymbol{Z})$ is:

\begin{align*}
&\exp\left(-\frac{1}2(\boldsymbol{y}-\boldsymbol{h}-\boldsymbol{X}\boldsymbol{\beta})'\boldsymbol{S}_{\boldsymbol{\gamma}}^{-1}(\boldsymbol{y}-\boldsymbol{h}-\boldsymbol{X}\boldsymbol{\beta}) - \frac{1}{2}\boldsymbol{h}'(\tau \boldsymbol{K}_{\boldsymbol{r}})^{-1}\boldsymbol{h}\right) \\
=&\exp\left(-\frac{1}2((\boldsymbol{y}-\boldsymbol{X}\boldsymbol{\beta}) -\boldsymbol{h})'\boldsymbol{S}_{\boldsymbol{\gamma}}^{-1}((\boldsymbol{y}-\boldsymbol{X}\boldsymbol{\beta}) -\boldsymbol{h}) - \frac{1}{2}\boldsymbol{h}'(\tau \boldsymbol{K}_{\boldsymbol{r}})^{-1}\boldsymbol{h}\right) \\
=&\exp\left(-\frac{1}2(\boldsymbol{y}-\boldsymbol{X}\boldsymbol{\beta})'\boldsymbol{S}_{\boldsymbol{\gamma}}^{-1}(\boldsymbol{y}-\boldsymbol{X}\boldsymbol{\beta}) + (\boldsymbol{y}-\boldsymbol{X\beta})'\boldsymbol{S}_{\boldsymbol{\gamma}}^{-1}\boldsymbol{h} -\frac{1}2\boldsymbol{h}'\boldsymbol{S}_{\boldsymbol{\gamma}}^{-1}\boldsymbol{h}-\frac{1}{2}\boldsymbol{h}'(\tau \boldsymbol{K}_{\boldsymbol{r}})^{-1}\boldsymbol{h}\right)\\
=&\exp\left(-\frac{1}2(\boldsymbol{y}-\boldsymbol{X}\boldsymbol{\beta})'\boldsymbol{S}_{\boldsymbol{\gamma}}^{-1}(\boldsymbol{y}-\boldsymbol{X}\boldsymbol{\beta}) + (\boldsymbol{y}-\boldsymbol{X\beta})'\boldsymbol{S}_{\boldsymbol{\gamma}}^{-1}\boldsymbol{h} -\frac{1}{2}\boldsymbol{h}'\left(\boldsymbol{S}_{\boldsymbol{\gamma}}^{-1} + (\tau\boldsymbol{K}_{\boldsymbol{r}})^{-1}\right)\boldsymbol{h}\right)
\end{align*}
and therefore:

\begin{align*}
&\int p(\boldsymbol{y} \mid \boldsymbol{\beta}, \boldsymbol{h}, \boldsymbol{\gamma}, \boldsymbol{X}, \boldsymbol{W})p(\boldsymbol{h}\mid \tau, \boldsymbol{r}, \boldsymbol{Z})d\boldsymbol{h}\\ 
\propto &\int \exp\left(-\frac{1}2(\boldsymbol{y}-\boldsymbol{X}\boldsymbol{\beta})'\boldsymbol{S}_{\boldsymbol{\gamma}}^{-1}(\boldsymbol{y}-\boldsymbol{X}\boldsymbol{\beta}) + (\boldsymbol{y}-\boldsymbol{X\beta})'\boldsymbol{S}_{\boldsymbol{\gamma}}^{-1}\boldsymbol{h} -\frac{1}{2}\boldsymbol{h}'\left(\boldsymbol{S}_{\boldsymbol{\gamma}}^{-1} + (\tau\boldsymbol{K}_{\boldsymbol{r}})^{-1}\right)\boldsymbol{h}\right)d\boldsymbol{h}\\
&\propto \exp\left(-\frac{1}2(\boldsymbol{y}-\boldsymbol{X}\boldsymbol{\beta})'\boldsymbol{S}_{\boldsymbol{\gamma}}^{-1}(\boldsymbol{y}-\boldsymbol{X}\boldsymbol{\beta})\right) \int  \exp\left((\boldsymbol{y}-\boldsymbol{X\beta})'\boldsymbol{S}_{\boldsymbol{\gamma}}^{-1}\boldsymbol{h} -\frac{1}{2}\boldsymbol{h}'\left(\boldsymbol{S}_{\boldsymbol{\gamma}}^{-1} + (\tau\boldsymbol{K}_{\boldsymbol{r}})^{-1}\right)\boldsymbol{h}\right)d\boldsymbol{h}
\end{align*}
Here, we recognize that the integral represents a multivariate normal moment-generating function for $\boldsymbol{h}$, up to a constant. To see this, we let $\Sigma^{-1} = \left(\boldsymbol{S}_{\boldsymbol{\gamma}}^{-1} + (\tau\boldsymbol{K}_{\boldsymbol{r}})^{-1}\right)$, $\boldsymbol{\mu} = \boldsymbol{0}$, and $\boldsymbol{t'} = (\boldsymbol{y}-\boldsymbol{X\beta})'\boldsymbol{S}_{\boldsymbol{\gamma}}^{-1}$. Using the Woodbury Matrix Identity, $\Sigma = \boldsymbol{S}_{\boldsymbol{\gamma}} - \boldsymbol{S}_{\boldsymbol{\gamma}}(\tau\boldsymbol{K}_{\boldsymbol{r}} + \boldsymbol{S}_{\boldsymbol{\gamma}})^{-1}\boldsymbol{S}_{\boldsymbol{\gamma}}$. This expression can therefore be re-written as:
\begin{align*}
&\propto \exp\left(-\frac{1}2(\boldsymbol{y}-\boldsymbol{X}\boldsymbol{\beta})'\boldsymbol{S}_{\boldsymbol{\gamma}}^{-1}(\boldsymbol{y}-\boldsymbol{X}\boldsymbol{\beta})\right) \exp\left(\frac{1}2(\boldsymbol{y}-\boldsymbol{X\beta})'\boldsymbol{S}_{\boldsymbol{\gamma}}^{-1}\left(\boldsymbol{S}_{\boldsymbol{\gamma}} - \boldsymbol{S}_{\boldsymbol{\gamma}}(\tau\boldsymbol{K}_{\boldsymbol{r}} + \boldsymbol{S}_{\boldsymbol{\gamma}})^{-1}\boldsymbol{S}_{\boldsymbol{\gamma}}\right)\boldsymbol{S}_{\boldsymbol{\gamma}}^{-1}(\boldsymbol{y}-\boldsymbol{X}\boldsymbol{\beta})\right) \\
&\propto \exp\left(-\frac{1}2(\boldsymbol{y}-\boldsymbol{X}\boldsymbol{\beta})'\boldsymbol{S}_{\boldsymbol{\gamma}}^{-1}(\boldsymbol{y}-\boldsymbol{X}\boldsymbol{\beta})\right) \exp\left(\frac{1}2(\boldsymbol{y}-\boldsymbol{X\beta})'\boldsymbol{S}_{\boldsymbol{\gamma}}^{-1}(\boldsymbol{y}-\boldsymbol{X\beta}) - \frac{1}2(\boldsymbol{y}-\boldsymbol{X\beta})'(\boldsymbol{S}_{\boldsymbol{\gamma}} + \tau\boldsymbol{K}_{\boldsymbol{r}})^{-1}(\boldsymbol{y}-\boldsymbol{X}\boldsymbol{\beta})\right) \\
&\propto\exp\left( - \frac{1}2(\boldsymbol{y}-\boldsymbol{X\beta})'(\boldsymbol{S}_{\boldsymbol{\gamma}} + \tau\boldsymbol{K}_{\boldsymbol{r}})^{-1}(\boldsymbol{y}-\boldsymbol{X}\boldsymbol{\beta})\right) 
\end{align*}
which is the kernel of a normal distribution with mean $\boldsymbol{X\beta}$ and covariance matrix $\boldsymbol{S}_{\boldsymbol{\gamma}} + \tau\boldsymbol{K}_{\boldsymbol{r}}$. So, 
\begin{align*}
\boldsymbol{y} \mid \boldsymbol{\beta}, \tau, \boldsymbol{r}, \boldsymbol{\gamma} \boldsymbol{X}, \boldsymbol{Z}, \boldsymbol{W} \sim MVN(\boldsymbol{\mu} = \boldsymbol{X\beta}, \boldsymbol{\Sigma} = \boldsymbol{S}_{\boldsymbol{\gamma}} + \tau\boldsymbol{K}_{\boldsymbol{r}}).
\end{align*}
After integrating out $\boldsymbol{h}$, the posterior distribution can be re-written as:
\begin{align*}
p(\boldsymbol{\beta}, \tau, \boldsymbol{r}, \boldsymbol{\gamma}\mid \boldsymbol{y}, \boldsymbol{X}, \boldsymbol{Z}, \boldsymbol{W}) 
&\propto p(\boldsymbol{y} \mid \boldsymbol{\beta}, \tau, \boldsymbol{r}, \boldsymbol{\gamma}, \boldsymbol{X}, \boldsymbol{Z}, \boldsymbol{W})p(\boldsymbol{\beta})p(\boldsymbol{r})p(\tau)p(\boldsymbol{\gamma}).
\end{align*}


\end{document}